\RequirePackage{fix-cm}
\documentclass[twocolumn,epjc3]{svjour3}  
\usepackage[utf8]{inputenc}
\usepackage{textgreek}
\smartqed  
\usepackage{color}
\usepackage{xcolor}
\definecolor{bred}{rgb}{0.8, 0.0, 0.0}
\definecolor{pblue}{rgb}{0.2, 0.2, 0.6}
\definecolor{ao}{rgb}{0.0, 0.5, 0.0}
\definecolor{carmine}{rgb}{0.59, 0.0, 0.09}

\newcommand{\keVee}{\,keV$_{ee}$}
\newcommand{\eVee}{\,eV$_{ee}$}

\newcommand{\CEvNS}{CE$\nu$NS}
\newcommand{\conusplus}{CONUS\texttt{+}}

\usepackage{graphicx}
\usepackage{amsmath}
\usepackage{amssymb}
\usepackage{hyperref}
\usepackage{cite}
\usepackage{upgreek}
\usepackage{subcaption}
\usepackage{url}
\usepackage{gensymb}
\usepackage{tabularx}
\usepackage{adjustbox}
\usepackage{arydshln}
\usepackage{rotating}

\usepackage{hyperref}
\hypersetup{
    colorlinks = true,
    linkbordercolor = {white},
    linkcolor = bred,
    citecolor = pblue,
    urlcolor = pblue
}

\usepackage[switch]{lineno} 

\begin{document}\sloppy

\title{Background decomposition of the \conusplus~run 1 data }

\author{N.~Ackermann\thanksref{e1,MPIK}, H.~Bonet\thanksref{MPIK}, C.~Buck\thanksref{MPIK}, J.~Hakenm\"{u}ller\thanksref{MPIK, Wien},  G.~Heusser\thanksref{MPIK}, M.~Lindner\thanksref{MPIK}, W.~Maneschg\thanksref{MPIK, canberra}, S.~Mertens\thanksref{MPIK}, K.~Ni\thanksref{MPIK}, D.~Piani\thanksref{MPIK}, M.~Rank\thanksref{KKL}, T.~Rink\thanksref{MPIK}, E.~S\'{a}nchez Garc\'{i}a\thanksref{MPIK}, I.~Stalder\thanksref{KKL}, H.~Strecker\thanksref{MPIK}, J.~Woenckhaus\thanksref{PSI}
}

\authorrunning{N. Ackermann et al.}
\institute{Max-Planck-Institut f\"ur Kernphysik, Saupfercheckweg 1, 69117 Heidelberg, Germany \label{MPIK} \and Kernkraftwerk Leibstadt AG, 5325 Leibstadt, Switzerland  \label{KKL} \and \emph{Present Address:}  Paul Scherrer Institut, Forschungsstrasse 111, 5232 Villigen, Switzerland\label{PSI} \and \emph{Present Address:} Marietta-Blau-Institut f\"ur Teilchenphysik der \"OAW, Dominikanerbastei 16, 1010 Wien, Austria\label{Wien} \and \emph{Present Address:} Mirion Technologies (Canberra) GmbH, Stahlstraße 42-44, 65428 Rüsselsheim, Germany\label{canberra} \vspace*{0.2cm} 
}
\thankstext{e1}{\href{mailto:ackerm@mpi-hd.mpg.de}{ackerm@mpi-hd.mpg.de (corresponding author)}}
\thankstext{e2}{\href{mailto:conus.eb@mpi-hd.mpg.de}{conus.eb@mpi-hd.mpg.de}}

\date{\today}

\maketitle

\begin{abstract}

The CONUS+ experiment is measuring the coherent elastic neutrino nucleus scattering (CE$\nu$NS) process using reactor anti-neutrinos as a source and four low energy threshold point-contact high-purity germanium spectrometers for their detection. It achieved the first measurement of coherent neutrino scattering at a nuclear reactor in run 1 of the experiment with a detection energy threshold of 160 eV$_{ee}$. This work presents the decomposition of the background spectra of the three detectors used in the run 1 analysis and the development of the corresponding background model with Geant4-based Monte Carlo simulations. The background model is used as the underlying input for the likelihood fit of the analysis. It is shown that reactor-correlated backgrounds are subdominant in all energy regions, specifically in the region of interest for CE$\nu$NS searches below 350 eV$_{ee}$ where their contribution is one order of magnitude below the expected CE$\nu$NS signal. Furthermore, cosmic ray muons and neutrons are identified as the dominant background source below 1 keV$_{ee}$ contributing approximately 75 - 90 \% of the recorded background rate. The final background model predicts an average rate of (47.5 $\pm$ 3.2) d$^{-1}$ kg$^{-1}$ in reactor on measurement in the energy region between [0.4, 1.0] keV$_{ee}$, which is in excellent agreement with the average measured value of (48.0 $\pm$ 0.6) d$^{-1}$ kg$^{-1}$. Similar agreement is found in all energy regions of both reactor on and off measurements.


\end{abstract}

\section{Introduction}
\label{intro}

The \conusplus~experiment is a low-background neutrino experiment measuring coherent elastic neutrino-nucleus scattering (\CEvNS) at the commercial nuclear power plant in Leibstadt, Switzerland (KKL). In run 1 of the experiment, which lasted from November 2023 to September 2024, the experiment employed four 1 kg high-purity germanium (HPGe) detectors which reached low-energy thresholds down to 0.16 \keVee (electron-equivalent). The experiment collected 327 kg d of reactor on and 60 kg d of reactor off data, resulting in the first successful measurement of the \CEvNS~interaction at reactor site with (395 $\pm$ 106) antineutrino events, corresponding to a rejection of the background-only hypothesis at 3.7$\sigma$ \cite{ackermann2025observationreactorantineutrinoscoherent}. During the measurement, the detectors are placed in a massive composite shield, effectively reducing background events in the germanium detectors by approximately four orders of magnitude and reaching levels of $\sim$70 d$^{-1}$ kg$^{-1}$ keV$^{-1}$ in the energy region of interest (ROI) for the \CEvNS~analysis between the detector thresholds and 800 \eVee, $\sim$20 d$^{-1}$ kg$^{-1}$ keV$^{-1}$ between 2 and 8 keV$_{ee}$, $\sim$5 d$^{-1}$ kg$^{-1}$ keV$^{-1}$ between 15 and 30 keV$_{ee}$, $\sim$4 d$^{-1}$ kg$^{-1}$ keV$^{-1}$ between 30 and 100 keV$_{ee}$, and $\sim$5 d$^{-1}$ kg$^{-1}$ keV$^{-1}$ between 100 and 250 keV$_{ee}$. Other similar CE$\nu$NS experiments located at nuclear reactors report similar or higher background levels close to their detection threshold: 62~d$^{-1}$~kg$^{-1}$~keV$^{-1}$ in the TEXONO experiment, 134~d$^{-1}$~kg$^{-1}$~keV$^{-1}$ in the NuGen experiment, and 3095~d$^{-1}$~kg$^{-1}$~keV$^{-1}$ in the DRESDEN-II experiment \cite{DresdenII_PRD_2021, DresdenII_PRL_2022, NuGEN_PRD_2022, TEXONO_TAUP_2023}. \newline 
The complete decomposition of the remaining measured background in all energy regions in both reactor on and reactor off measurements for CONUS+ was achieved using Monte Carlo simulations. The general approach is explained in Section~\ref{sec:sim_frame}. The simulations are based on the framework set up for the analogue study performed already at the CONUS experiment site at KBR \cite{conus_bkg} and on the experimental results obtained from a background characterization campaign at the CONUS+ location at KKL prior to detector installation \cite{conusplus_bkg} (see Section~\ref{sec:back_char}). \newline
All possible background sources for the CONUS+ experiment were investigated. Section~\ref{sec:reactorcorr} describes the simulation of reactor-correlated background components, namely neutrons and inert gases. Reactor neutrons can be critical for a CE$\nu$NS experiment, since they can mimic a possible CE$\nu$NS signature in the germanium detectors and are only present during reactor on measurements. Section~\ref{sec:cosmic} investigates the influence of cosmic ray muons and neutrons, the dominant source of background in the low energy spectrum of CONUS+. Section~\ref{sec:radon} deals with air-borne radon in the detector chamber which can diffuse through the shield from the outside air and has a large impact on the high energy part of the CONUS+ data. Another source of background are intrinsic or cosmogenically activated contaminations in the materials of the setup and the germanium crystals themselves, which are described in Section~\ref{sec:activation}. Section~\ref{sec:additional} summarizes the remaining smaller background components, which lastly leads to the full decomposition of the background of the CONUS+ run 1 data in both reactor on and off measurements presented in Section~\ref{sec:fullmodel}. \newline
This full model is a crucial input for the successful first measurement of CE$\nu$NS at a nuclear reactor by CONUS+. As described in \cite{ackermann2025observationreactorantineutrinoscoherent}, the analysis of the data relies on a likelihood fit where a model of the reactor on data, consisting of the background model presented here and the CE$\nu$NS signal prediction, is fitted to the data. The correctness of the background model especially in the ROI is therefore crucial to ensure a correct fit result and to determine the number of CE$\nu$NS events in the measured reactor on data.  

\section{The CONUS+ experiment}
\label{sec:exp_input}

\subsection{Experimental location}
\label{sec:back_char}

\begin{figure}
    \centering
    \includegraphics[width=0.48\textwidth]{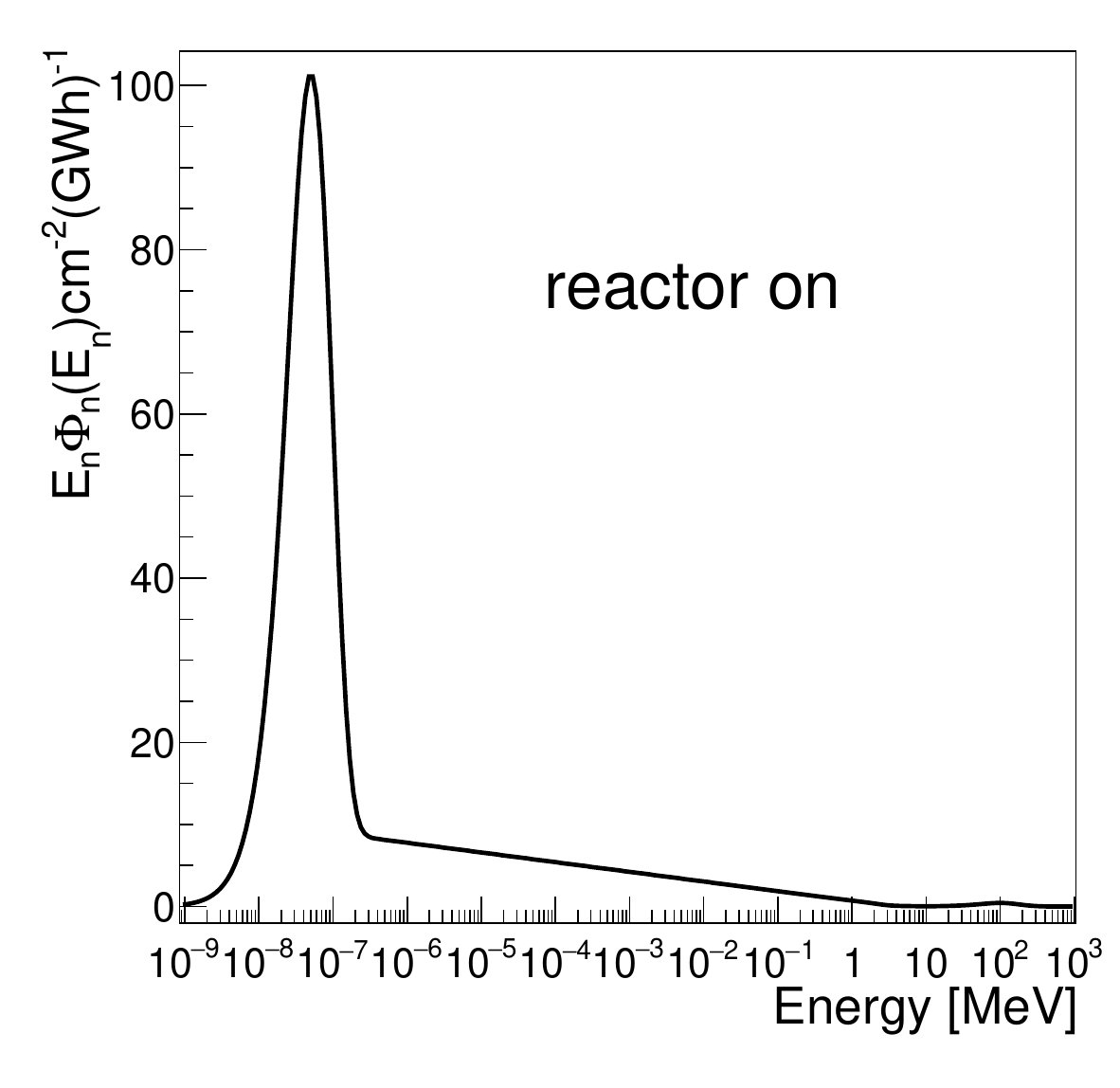}
    \caption{Measured neutron energy distribution $\phi_{on}$(E$_{n})$ resulting from the analysis of the reactor on data normalized to the energy emitted by the reactor, as published in \cite{conusplus_bkg}.}
    \label{fig:spectra_ON}
\end{figure}

The CONUS+ experiment is located in the reactor building of the KKL powerplant in room ZA28R027 at a distance of 20.7 m from the reactor core.  The four HPGe detectors C2, C3, C4, and C5 are placed in a massive shield (see Section~\ref{sec:shield}) to mitigate the effect of background sources on the measurement. The shield is surrounded by the concrete walls of room ZA28R027, which itself is surrounded by the outer walls of the reactor building consisting of 3.8 cm of steel (steel containment) and 1.2 m of reinforced concrete. Details on the densities of these materials can be found in Section~\ref{sec:cosmicn}. In total these layers add up to an effective overburden of (7.3 $\pm$ 0.1) meters in water equivalent (m w.e.) effectively reducing the expected hadronic part of the cosmic ray flux by two orders of magnitude in the room. Additionally, the many layers of concrete between the experimental room and the reactor core reduce radiation from the reactor in the room by several orders of magnitude to a level below 1 $\mu$Sv/h. The KKL reactor is regularly shut off for a period of one month per year to allow for maintenance work. During this outage the reactor core is opened and the top part of the reactor vessel, called the drywell lid, is removed and placed directly on top of the ZA28R027 room. It is made from 3.6 cm of steel and therefore increases the overburden by approximately 0.3 m w.e. This leads to a reduced flux of cosmic rays in the room during the outage. \newline
Before the installation of the CONUS+ experiment at KKL, an extensive background characterization campaign was carried out to assess the situation at the experimental site. The results are detailed in \cite{conusplus_bkg}. For gamma radiation in the room, the measurement campaign found a reduced integral rate above 2.7 MeV (above contributions from natural radioactivity) compared to the situation at the KBR power plant in Brokdorf, Germany.  A rate of 6 counts s$^{-1}$ kg$^{-1}$ was measured, with the remaining radiation coming from the capture of reactor neutrons on nuclei of the surrounding materials, mainly silicon and calcium isotopes in the concrete of the walls. \newline
The muon flux in the CONUS+ room was measured as ($107 \pm 3$) muons s$^{-1}$ m$^{-2}$, which is 1.9 times smaller than the flux at surface (without overburden). During reactor outage, it is found that the muon flux in the room is reduced by approximately 3 $\%$, which is consistent with the placement of the drywell lid above the CONUS+ room. \newline 
The neutron flux in the room during normal reactor operation was also measured. The overall reactor neutron fluence is 30 times higher than that for CONUS at KBR, but the impact of reactor neutrons on the overall background of the experiment is still expected to be small due to the excellent shielding capabilities of the CONUS+ shield, as will be shown in Section~\ref{sec:reactorcorr}. The resulting spectrum is shown in Figure~\ref{fig:spectra_ON} and will be used as input for the reactor neutron simulations presented in Section~\ref{sec:reactorcorr}.

In the neutron measurement during the reactor outage, a significant decrease in neutron flux was observed, as expected from the absence of reactor neutrons in this measurement. The remaining neutrons originate from cosmic rays and are an important source of background for the experiment, as detailed in Section~\ref{sec:cosmicn}. Due to constraints detailed in \cite{conusplus_bkg}, no reliable cosmic neutron spectrum could be extracted from the measurement. The corresponding simulations in \ref{sec:cosmicn} will instead be based on simulations of the propagation of the surface neutron flux through the reactor building.

\subsection{CONUS+ shield}
\label{sec:shield}

\begin{figure}
    \centering
    \includegraphics[width=0.48\textwidth]{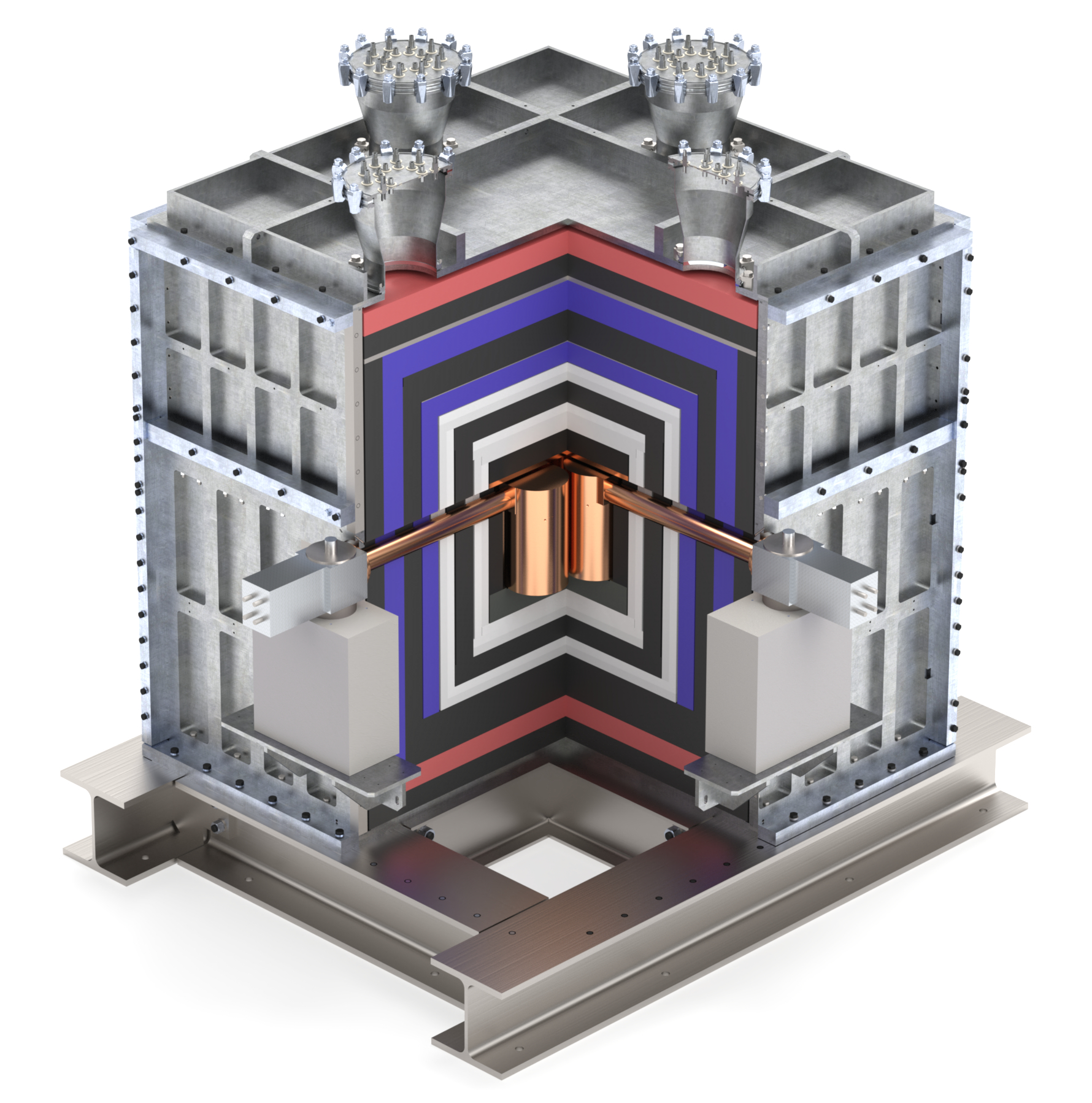}
    \caption{Drawing of the CONUS+ shield. The shield consists of several lead layers (black), borated polyethylene (PE) layers (white), two layers of plastic scintillator used as muon veto systems (blue), and unborated PE (red). The four detectors sit in the centre of the shield and the detector chamber is continuously flushed with radon-free air. The whole structure is encased in a steel frame for structural integrity.}
    \label{fig:shield}
\end{figure}

The shield setup of the CONUS+ experiment can be seen in Figure~\ref{fig:shield}. The design is largely based on the shield of the previous CONUS experiment at the KBR power plant in Brokdorf, Germany. Details on the overall structure and setup can be found in \cite{Hakenm_ller_2019, conusplus} . \newline
The passive parts of the shield (lead, (borated) PE) and the KBR plastic scintillator muon veto are recycled from the CONUS setup. An important change in the new setup is the replacement of one of the outer lead layers with an additional plastic scintillator layer acting as a second inner muon veto. The change was mainly motivated by the higher muon flux in the CONUS+ experiment compared to the CONUS experiment due to the smaller overburden of only ($7.3 \pm 0.1$) m w.e. The resulting smaller lead shield thickness at the CONUS+ shield was acceptable due to the smaller flux of high energy gamma radiation in the environment (see \ref{sec:back_char}). The new anti-coincidence muon veto system now consists of 18 plastic scintillator plates with a thickness of 5 cm which are arranged in two layers (an inner and outer veto) in the shield. Each plate is equipped with two PMTs (four in the top plate). The PMTs and plastic scintillator plates were previously screened at the Low Level Laboratory at MPIK and the results can be found in Table~\ref{tab:vetocomponents}. Dedicated simulations showed negligible impact from these contaminations in the ROI of CONUS+.

\begin{table}[bht]
\caption{Radiopurity of components of inner muon veto measured at MPIK}
\label{tab:vetocomponents}
\centering
\setlength\extrarowheight{4pt}
\begin{tabular}{lcc}
\hline
Component & Isotope & Activity [mBq/kg]\\
\hline
Inner muon veto plates & $^{226}$Ra & 0.09 $\pm$ 0.05 \\
 & $^{228}$Th chain & 0.5 $\pm$ 0.1 \\
 & $^{40}$K & 1.5 $\pm$ 0.5 \\
 & $^{228}$Ra chain& 0.4 $\pm$ 0.1 \\
\hline
Inner muon veto PMTs & $^{226}$Ra & 12.9 $\pm$ 2.5 \\
 & $^{228}$Th chain & 13 $\pm$ 3 \\
 & $^{40}$K & $<$ 221  \\
 & $^{228}$Ra chain& $<$ 15 \\
\hline
\end{tabular}
\end{table}

\subsection{Data collection}
\label{sec:data_collection}

The background model presented in this work describes the data collected in run 1 of the CONUS+ experiment. The CE$\nu$NS analysis in \cite{ackermann2025observationreactorantineutrinoscoherent} only includes data from three of the four detectors of run 1, namely C2, C3, and C5. The C4 detector is excluded for the reasons listed in \cite{ackermann2025observationreactorantineutrinoscoherent}. 
Before the start of the measurement campaign, the detectors were sent to Mirion Lingolsheim, where they underwent several upgrades detailed in \cite{conusplus}. New components which were added or replaced in the cryostat and electronics of the detectors during the upgrade were previously screened at MPIK with the screening stations of the Low Level Laboratory \cite{heusser2015giove}. Specifically the solder, flex and coax cables, PCB and SMD parts were investigated and the screening results are listed in Table \ref{tab:screening}. It was determined that all components meet the desired levels of radiopurity. Of particular note is the $^{210}$Pb level of the solder that is used for the installation of the surface-mount devices (SMD parts) on the printed circuit board (PCB) at 80 Bq/kg. Although this is by far the largest value of the measured isotopes in all components, the amount of solder that is needed for the upgrade is very low and thus is of little consequence for the background of the detector.  \newline

\begin{table*}[ht]
\caption{Screening results of new electronic parts for detector upgrade
\label{tab:screening}}
\begin{tabular}{lccccc}
\hline\noalign{\smallskip}
   & Solder    &   Flex cable & Coax cable& PCB (Kapton)& SMD parts \\\hline
Mass  & $\approx$ 1 g & $\approx$ 25 g  & $\approx$ 30 g  & $\approx$ 1 g  &  $<$ 1g  \\
$^{210}$Pb  & 80 Bq/kg      &    - & - &  - & - \\
$^{234m}$Pa  &  -  & $<$ 43 mBq/kg    & $<$ 95 mBq/kg    &$<$ 507 mBq/kg    & $<$ 1738 mBq/kg \\
$^{226}$Ra  &  -  & $<$ 24 mBq/kg  & $<$ 373 mBq/kg  & $<$ 366 mBq/kg  & 8049 $\pm$ 570 mBq/kg \\
$^{228}$Th  &  -  &  $<$ 48 mBq/kg  & $<$ 393 mBq/kg  &$<$ 739 mBq/kg  & 3208 $\pm$ 598 mBq/kg \\
$^{228}$Ra  &  -  & $<$ 101 mBq/kg  & $<$ 792 mBq/kg  & $<$ 827 mBq/kg  & $<$ 1480 mBq/kg \\
$^{137}$Cs  &  -  & $<$ 20 mBq/kg  & $<$ 113 mBq/kg  &$<$ 200 mBq/kg  & $<$ 151 mBq/kg  \\
$^{60}$Co   & -  & $<$ 16 mBq/kg  & $<$ 165 mBq/kg  &$<$ 214 mBq/kg  & $<$ 195 mBq/kg \\
$^{40}$K     &  -  &  $<$ 128 mBq/kg & $<$ 3029 mBq/kg &$<$ 3963 mBq/kg  & $<$ 1582 mBq/kg  \\      
\noalign{\smallskip}\hline
\end{tabular}
\end{table*}

In September 2023, the detectors were brought to KKL, where they were mounted in the shield inside of the containment area of the power plant. Physics data taking for run 1 of the experiment started in November of 2023 and ran continuously until July 2024, including a period of approximately one month in May 2024, when the reactor of the KKL power plant was shut down because of a routine maintenance outage. The data collected during this outage is a valuable input for the background modeling, since it is used as a confirmation of the model in absence of a neutrino signal (see  Section~\ref{sec:fullmodel}). The environmental conditions during an outage at KKL, however, are slightly changed compared to reactor on data collection periods, since the drywell lid of the reactor vessel, which is removed by KKL personnel for access to the reactor, is placed directly above the CONUS+ experimental room. This drywell lid, which is made of 3.6 cm of steel, therefore changes the overburden of the experiment and influences the suppression of cosmic muon and neutron fluxes in the room itself. Additional smaller changes in the background model in the reactor off period come from inert gases, which are released when the reactor vessel is opened. \newline
The CONUS+ experiment takes data in two separate configurations for each of its four detectors: a low energy channel ranging from 0 to 35 keV$_{ee}$, which includes the ROI, and a high energy channel which ranges up to several hundred keV$_{ee}$. The high energy channel is only used for the construction of the background model. The difference in the configurations comes from a selected higher gain in the low E channels at DAQ level, allowing for detailed studies in the low energy channels while the high energy channels are used as input for the construction of the background model. \newline
In the measurements for run 1 of the CONUS+ experiments, it was noticed that the high energy channels for all detectors feature large efficiency losses. The cause of these efficiency losses is currently under study, but as of the final analysis of the run 1 data no clear reason could be identified. In order to still be able to use the high energy channels for the background model, the efficiency loss is corrected using the results of muon simulations. The procedure will be detailed in Section~\ref{highEcorrection}. The method, while producing workable spectra, introduces large uncertainties in the count rates for bins over 300 keV$_{ee}$ due to the limited statistics in these bins in the original data sets. As a result, only
energies up to 300 keV$_{ee}$ are used for analysis and information about background sources.

\section{Simulation framework and approach}
\label{sec:sim_frame}

The Monte Carlo (MC) simulations performed in the context of the CONUS+ experiment use the MaGe \cite{mage} framework introduced for the Majorana and Gerda experiments \cite{majorana, gerda}. This framework is based on Geant4 (version 10.4.3) and is specifically adapted to be used for low energy experiments relying on a list of specific physics processes relevant in this regime. It has been thoroughly validated in the past \cite{GERDA_LArVeto_2023, Majorana_2019, LEGEND_1000_2021, HPGe_Screening_2023} in other experiments, as well as in the context of the CONUS experiment at KBR for energies below 10 keV$_{ee}$ \cite{conus_bkg}. In the framework, the geometry of the CONUS+ detectors including the copper cryostats, as well as the full shield configuration, and the reactor building of KKL are implemented with their dimensions, materials, and respective densities taken from technical drawings. Within MaGe, the physics lists chosen for the simulations are the "Livermore" physics model for low energy electromagnetic processes, as well as the "Neutron High Precision model" (NeutronHP) for accurate neutron propagation below energies of 20 MeV. Furthermore, the "QGSP-BERT-HP" model is used for the simulations presented in this work with the "Bertini" cascade model for nucleon and pion interactions below 10 GeV \cite{Heikkinen:2003wd}. Production cuts for secondary particles are set to the "DarkMatter" option, the lowest energy realm offered in MaGe. \newline 
The simulation results are processed with a dedicated post-processing routine designed for the CONUS+ experiment. In this routine, in a first step a quenching factor is applied to all energy depositions by a hadron (mostly neutrons). The quenching factor is calculated from the Lindhard model \cite{lindhard1963range}, which has been well verified in the context of the CONUS experiment through a dedicated measurement campaign, as well as the previously published CE$\nu$NS results \cite{Bonhomme_2022, Ackermann_2024, ackermann2025observationreactorantineutrinoscoherent}. Next, an energy detection efficiency is applied to all energy depositions. This efficiency describes the three different typical zones that can be found in a HPGe detector: the dead layer on the very outside, where detection efficiency is zero, the transition layer, where the detection efficiency is described with a rising sigmoid function towards the centre of the detector, and the bulk volume where detection efficiency is one \cite{conus_bkg}. To determine the position of the energy deposition in the crystal, the coordinates from the Geant4 simulation are used. Lastly, all energy depositions of one primary particle are summed to obtain the final measurable energy in the detector. In a final step, the energy resolution of the detector is applied to the simulation result by convolution of the MC spectrum with a Gaussian. The energy-dependent width of the Gaussian is given by three parts: the pulser resolution of each detector, which is measured independently and monitored during the data collection period, and the inherent statistical fluctuation of the number of charge carriers created in an interaction inside the germanium, which is given by the Fano factor, and the energy it takes to create an electron-hole pair in germanium. This energy is 2.96 eV, while the Fano factor is assumed to be 0.11. The third contributing part corresponds to the charge carrier collection efficiency. This depends on the concentration of vacancies in the bulk of the Ge diodes. \newline
The general approach in the building of the background model is focused on blindness in the energy region in which the CE$\nu$NS signal is expected to appear (from detector threshold to 400 eV$_{ee}$ in the reactor on data). As such, only information from energy ranges above 400 eV$_{ee}$ in the reactor on data and the full reactor off data are used to validate and normalize the different background components which are presented in the following sections. The energy range below 400 eV$_{ee}$ is only unblinded in the final likelihood fit of the analysis to avoid any bias in the creation of the background model.

\section{Reactor-induced backgrounds}
\label{sec:reactorcorr}

The first source of background being investigated in the CONUS+ experiment are backgrounds directly induced by the reactor. These can either originate from ongoing reactor operation, like reactor thermal power correlated neutrons (short: reactor neutrons), or be released when the reactor vessel is opened, like certain inert gases. The following section will analyse these contributions. 

\subsection{Reactor neutrons}
\label{sec:reactorneutrons}

\begin{figure}[ht]
    \centering
    \includegraphics[width=0.47\textwidth]{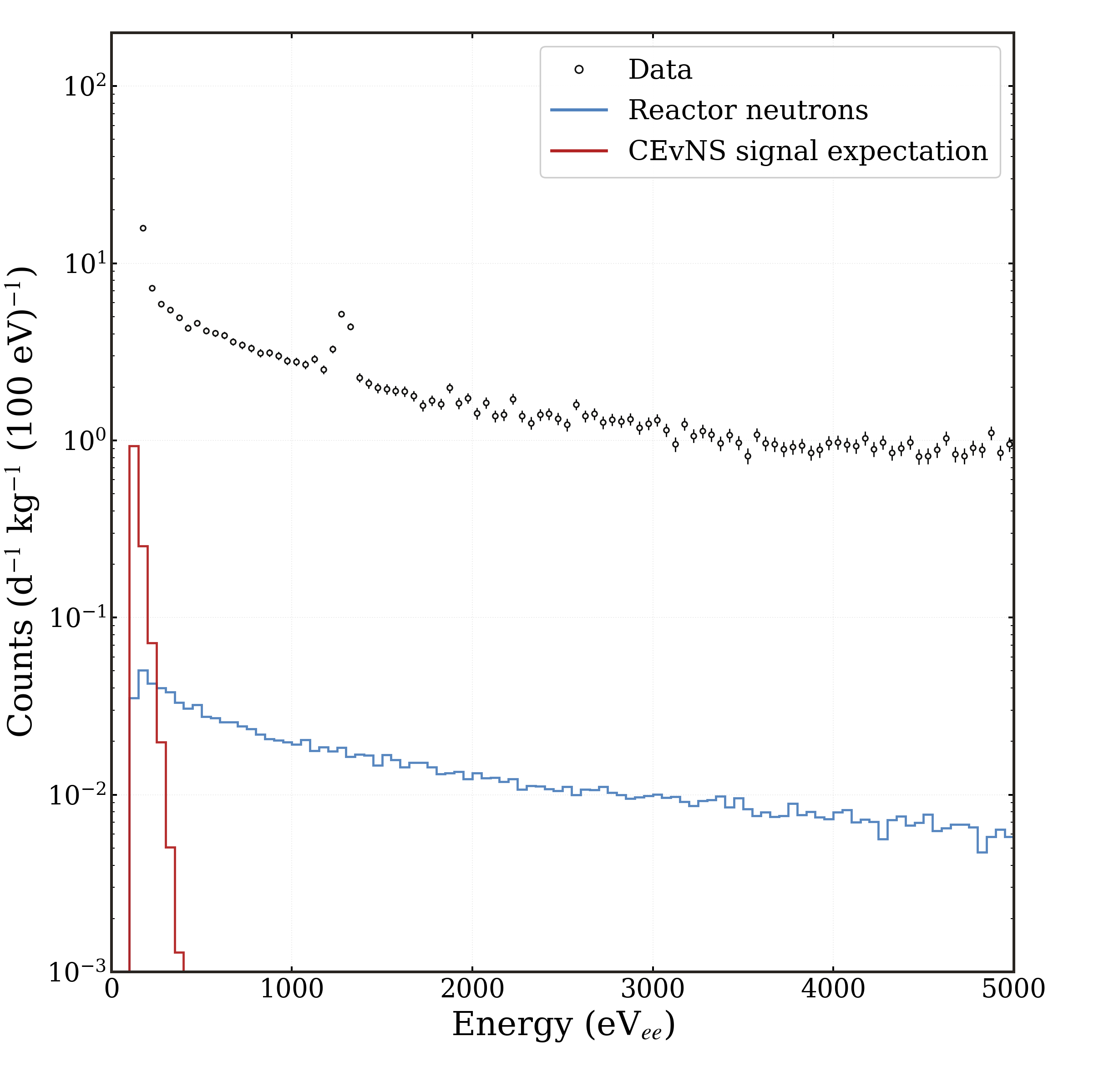}
    \caption{Simulation result of the reactor neutrons compared to the C5 spectrum during reactor on measurement. The simulation is also compared to the expected CE$\nu$NS signal in the C5 detector.}
    \label{fig:reactor_neutrons}
\end{figure}

As described in Section~\ref{sec:back_char}, a flux of neutrons arrives in the CONUS+ room from fission reactions in the reactor core. These neutrons can be especially problematic for CE$\nu$NS searches, since they can also scatter off the germanium nuclei in the crystals, both elastically and inelastically, and their flux is correlated to the thermal power of the reactor \cite{Hakenm_ller_2019}. They can therefore mimic the expected CE$\nu$NS signature. Figure~\ref{fig:reactor_neutrons} shows the measured spectrum of neutrons in the room during active reactor operation. From the small count rates in all Bonner spheres during the reactor off measurement, it can be assumed that nearly all neutrons in Figure~\ref{fig:reactor_neutrons} originate from the reactor. To assess their impact on the background, this spectrum (up to 20 MeV, no cosmic neutrons) is used as input for a simulation where 10$^{10}$ neutrons are started isotropically from a half-sphere around the CONUS+ shield with a radius of 1.5 m.  Figure~\ref{fig:reactor_neutrons} shows the result of this simulation compared to C5 run 1 data.

The reactor neutron background contribution shows  a steep increase towards low energies. This general shape will also be seen for the impact of cosmic neutrons in Section \ref{sec:cosmicn} However, their overall contribution to the background is only (0.2 $\pm$ 0.1) counts d$^{-1}$ kg$^{-1}$ between 0.4 and 1 keV$_{ee}$, which is around two orders of magnitude smaller than that of the cosmic neutrons and below 1 $\%$ of the total background in that energy region. Their impact is still around 20 times higher than what was found for the CONUS experiment in KBR, which is consistent with the increased neutron flux in the room. As reactor neutrons can mimic the appearance of a CE$\nu$NS signal in the detector, the comparison of this component to the expected signal in Figure~\ref{fig:reactor_neutrons} is crucial. It can be seen that the expected signal is around one order of magnitude bigger than the impact of the reactor neutrons below 300 eV$_{ee}$, which rules out the possibility of a wrong identification in the analysis presented in \cite{ackermann2025observationreactorantineutrinoscoherent}. 

\subsection{Inert gases}
\label{sec:inertgases}

Radioactive isotopes of inert gases are continuously produced in the reactor. They can escape when the reactor is opened in the maintenance outage period as well as during tests of the Reactor Core Isolation Cooling (RCIC) system which are performed approximately every two months. Similarly to radon, they can then diffuse into the CONUS+ shield, where they can decay and induce backgrounds. The most important of these isotopes for CONUS+ are $^{3}$H, $^{85}$Kr, and $^{135}$Xe.  $^{85}$Kr and $^{135}$Xe can be produced as direct fission products in the core or from the decay of $^{135}$I, in the case of $^{135}$Xe. $^{3}$H is primarily produced from the interaction of neutrons with boron used as a neutron control material in the reactor. $^{3}$H and $^{85}$Kr are relatively long-lived isotopes with half-lives of 12 and 11 years respectively, while $^{135}$Xe has a short half-life of only 9 hours. Its impact on the CONUS+ background is therefore confined to the first few days after the opening of the reactor vessel, while $^{3}$H and $^{85}$Kr can have a longer effect. The decay spectra of the latter two also do not feature a gamma-ray peak in the sensitive range of the CONUS+ experiment, while the decay of $^{135}$Xe to $^{135}$Cs induces a gamma-ray with an energy of 250 keV. This peak is visible in the run 1 data of all detectors, primarily in the reactor off spectra. \newline
For the simulations, the gas was placed homogeneously in the inner detector chamber of the setup. The result of the xenon simulation is scaled to match the count rate of the 250 keV peak in the measured spectra. The contributions from $^{3}$H and $^{85}$Kr were estimated based on the known yield of the separate isotopes per fission compared to that of $^{135}$Xe, taken from evaluated nuclear data libraries (e.g. ENDF/B-VIII.0 \cite{Brown2018ENDF}). The resulting background spectra are shown alongside the full model in Figures \ref{fig:fullmodel_c5} - \ref{fig:fullmodel_c3}. In total, inert gases from the reactor contribute around 1 to 3 $\%$ of the total background of the detectors between 400 and 1000 eV$_{ee}$, depending on the detector and reactor on or off measurement.

\section{Cosmic ray background}
\label{sec:cosmic}

The next source of background being investigated for the CONUS+ experiment are cosmic rays. It was already shown for the CONUS experiment at KBR that muons are one of the dominant sources of background for experiments of this type \cite{conus_bkg}. This is exacerbated by the fact that the CONUS+ experiment only features an effective overburden of (7.3 $\pm$ 0.1) m w.e. compared to the 24 m w.e. found for CONUS, which leads to an increase of the muon flux in the room by a factor of 2.5. A second consequence of the lowered overburden is the presence of a small amount of high energy cosmic neutrons (E $\sim$ 100 MeV) in the room, which are only effectively shielded at overburdens larger than 10 m w.e \cite{Heusser:1995wd}.  For an overburden of 7.3 m w.e. a reduction by two orders of magnitude is expected. These neutrons with energies of around 100 MeV can penetrate the CONUS+ shield and also induce background in the germanium diodes.

\subsection{Cosmic ray muons}
\label{sec:cosmicmu}

\begin{figure}
    \centering
    \includegraphics[width=0.48\textwidth]{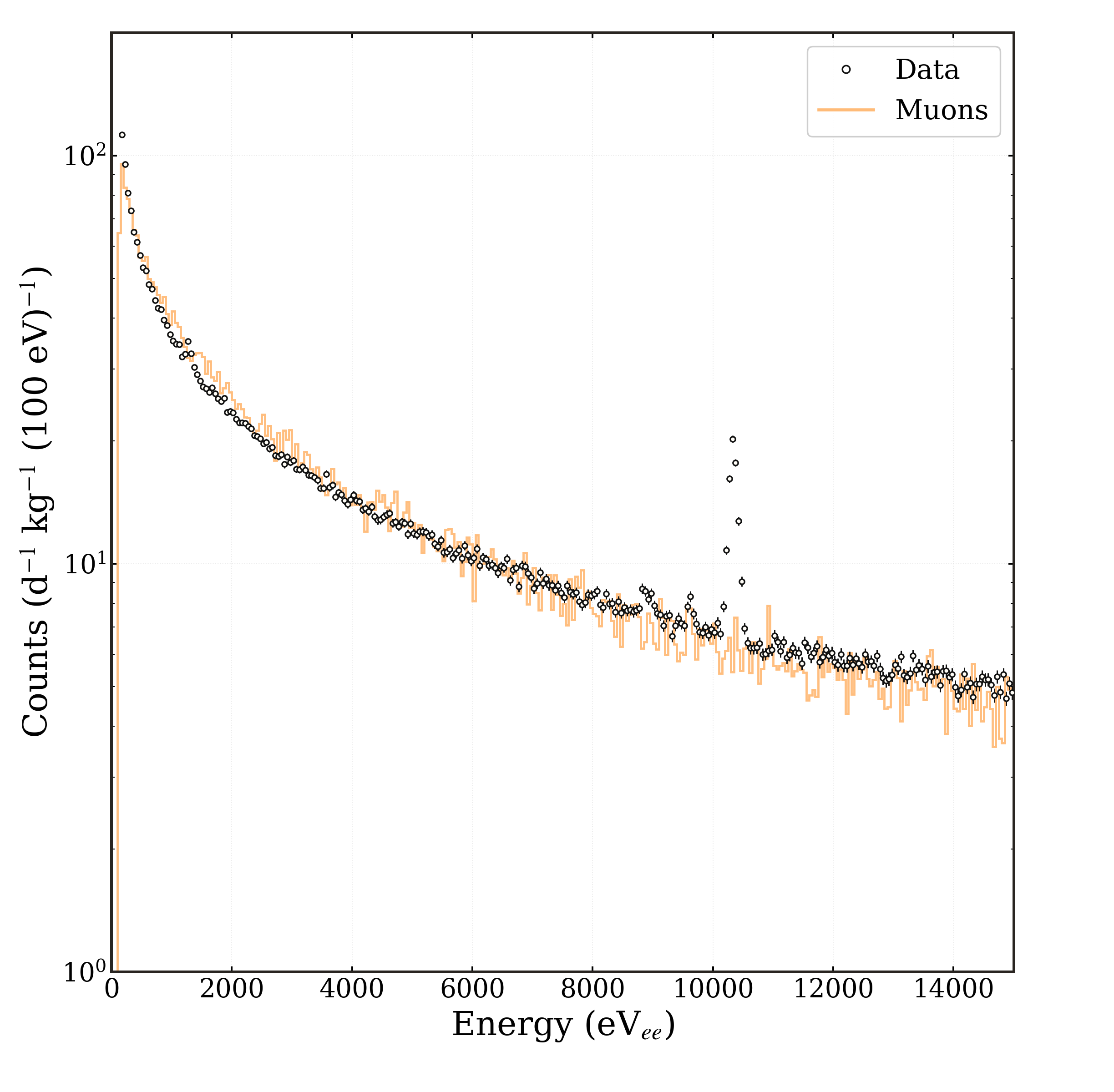}
    \caption{Comparison of the total simulated muon background spectrum and data taken by the C5 detector in run 1 without the application of the muon veto from 0.4 to 30 keV$_{ee}$. Very good agreement is found. At 10.3 keV the gamma-ray line from the decay of $^{68/71}$Ge can be seen in the data.}
    \label{fig:muon_comp_noVeto}
\end{figure}

Cosmic muons, produced in the upper atmosphere primarily from pion and kaon decays, penetrate the overburden of the CONUS+ experiment and can induce background events in the germanium detectors through a number of different processes. These include the production of neutrons through muon-induced spallation, photonuclear reactions or muon capture when travelling through any material, specifically through the high-Z lead layers of the CONUS+ shield. Additionally, they can induce gamma radiation through bremsstrahlung and secondary electromagnetic showers. \newline
The flux, energies, and angular distributions of cosmic muons in the CONUS+ room are calculated from models for shallow underground laboratories in the literature \cite{reyna, Bugaev} as was previously validated in \cite{conus_bkg}. From the overburden a flux of  ($60.6 \pm 3.6$) $s^{-1}$ $m^{-2}$ is calculated. In the MC simulations, 10$^9$ muons and antimuons are started from each wall and the ceiling of the CONUS+ experimental room. In the postprocessing stage of the simulations, the muon charge ratio is considered in the normalisation of the results, which is taken to be 1.2776 \cite{CMS:2010qjf}. The results can be verified directly by comparing the MC simulation results to the CONUS+ spectra taken without the muon veto applied, which are dominated by muon-induced background. As seen in Figure~\ref{fig:muon_comp_noVeto}, the simulations give very good agreement with the data. 

As mentioned in Section~\ref{sec:shield}, the CONUS+ shield features two plastic scintillator layers which function as an inner and outer muon veto for the experiment. Whenever a muon travels through the plastic scintillator it deposits a characteristic amount of energy of approximately 2 MeV per cm. The PMTs in the corners of the veto plates collect the produced light and trigger whenever energy above a predefined threshold is deposited. In the offline analysis of the data, the time after such a hit is then cut with a "muon veto window" of 450 $\mu$s. The length of the window is chosen to also account for secondary neutrons produced by muons passing through the  shield. The inclusion of the muon veto system strongly suppresses the otherwise dominant muon background and efficiencies of up to 99 $\%$ are reached in the rejection of muon events.  In order to reproduce the veto cut in the MC simulations, in a first step the efficiency of the veto system is modelled by applying a constant reduction factor to the MC simulation results in Figure~\ref{fig:muon_comp_noVeto}. A first estimation of this value can be extracted from the overall background reduction that is achieved with the muon veto cut. 

\begin{figure}
    \centering
    \includegraphics[width=0.48\textwidth]{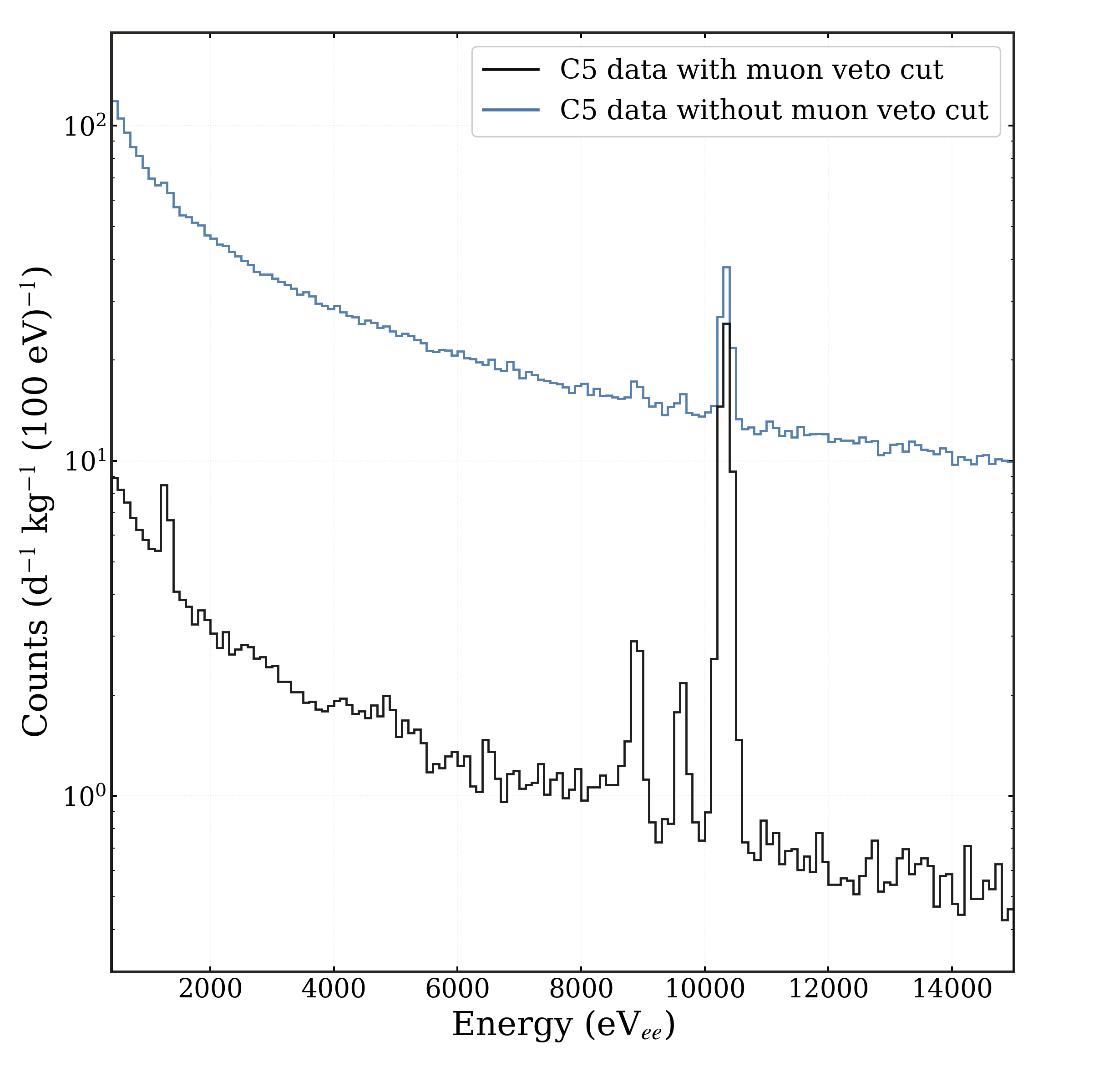}
    \caption{Comparison of C5 run 1 data with and without the applied muon veto cut.}
    \label{fig:muon_wandwoVeto}
\end{figure}

Figure~\ref{fig:muon_wandwoVeto} shows a comparison of data taken in run 1 of the CONUS+ experiment with and without applied muon veto cut. A large reduction of the event rate can be observed, which is consistent with the fact that the data without muon veto cut is completely dominated by muon-induced events. Table~\ref{tab::bkg_red_muon_cut} shows that a consistent background reduction of over 92 $\%$ is achieved in all energy regions of the data. The reduction is larger in higher energy regions indicating a larger influence of muon-induced events in these regions. However, it must be noted that the background reduction achieved with the cut cannot directly be used as cut efficiency in the muon simulations. This efficiency must be higher than 92 $\%$ due to the presence of other backgrounds in the data. \newline

\begin{table}[bht]
\caption{Background reduction in the C5 detector with the muon veto cut in different energy regions. }
\label{tab::bkg_red_muon_cut}
\centering
\small
\setlength{\tabcolsep}{4pt} 
\setlength{\extrarowheight}{2pt}

\begin{tabular}{lccc}
\hline
Energy region &
\begin{tabular}[c]{@{}c@{}}Count rate\\ without veto\\ {[}d$^{-1}$\,kg$^{-1}${]}\end{tabular} &
\begin{tabular}[c]{@{}c@{}}Count rate\\ with veto\\ {[}d$^{-1}$\,kg$^{-1}${]}\end{tabular} &
Ratio (\%) \\
\hline
(0.4--1) keV$_{ee}$  & 562 $\pm$ 2  & 43 $\pm$ 1  & 7.7 \\
(2--8) keV$_{ee}$    & 1575 $\pm$ 3 & 103 $\pm$ 1 & 6.5 \\
(15--30) keV$_{ee}$  & 1106 $\pm$ 3 & 53 $\pm$ 1  & 4.9 \\
\hline
\end{tabular}
\end{table}

\begin{figure}[bht]
    \centering
    \includegraphics[width=0.48\textwidth]{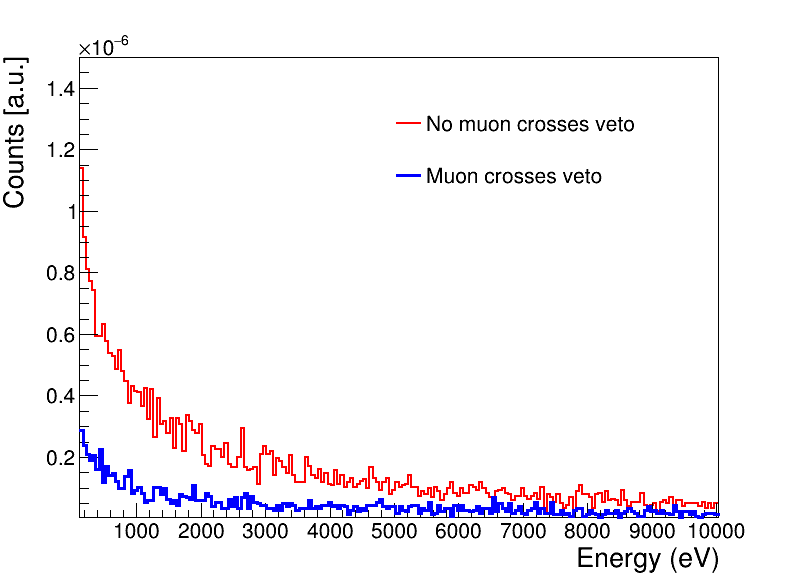}
    \caption{Simulated spectrum of muon-induced background events in the CONUS+ detectors from muons that do not cross a muon veto layer (red) and from muons that cross a muon veto layer (blue). At energies below 15 keV$_{ee}$, it can be seen that most of the events originate from muons that do not cross a plastic scintillator plate. These events are not triggered in the veto system and therefore induce an inefficiency at low energies.}
    \label{fig:muon_tag_wandwoveto}
\end{figure}

For the overall muon veto efficiency, a value of 99.0 $\%$ is found for all energies above 15 keV$_{ee}$. This is validated by the overall agreement of the background model with the data in all detectors and measurement periods (see Section~\ref{sec:fullmodel}), as well as by dedicated studies of the muon veto system prior to the installation of the experiment at KKL. At energies below 15 keV$_{ee}$ an additional effect leads to corrections in the muon veto efficiency. As shown in Figure~\ref{fig:shield}, both plastic scintillator layers in the shielding setup are surrounded by lead. This setup was chosen to reduce the impact of high energy gamma radiation on the muon veto system which would lead to the accumulation of a large amount of dead time \cite{conusplus}. However, due to this design, it is possible that an incoming muon passes through the outside layers of the shield without ever passing through one of the scintillator plates. Such a muon can then still induce electromagnetic showers in the shield which can lead to background in the germanium detectors. If these showers do not deposit enough energy in the veto system, the system will not trigger. The setup of the shield thus leads to an inefficiency in the tagging ability of the muon veto. The effect was studied in detail in MC simulations by performing the aforementioned muon simulations and additionally tagging all events where a muon never crosses any scintillator plates. The results of this study in terms of energy deposited in the germanium crystal can be found in Figure~\ref{fig:muon_tag_wandwoveto}. 

It can be seen that at very low energies, a majority of the energy deposited in the germanium crystals originates from events where no muon crosses the veto system. The effect generally increases towards lower energies such that in the critical energy region from 160 eV$_{ee}$ to 400 eV$_{ee}$, approximately 81 $\%$ of muon-induced energy depositions arise from this effect. In order to quantify the effect in a simplified way and convert it into a factor that can be applied to the MC simulations, the ratio of the two plots in Figure~\ref{fig:muon_tag_wandwoveto} is studied. Computing this ratio between the spectrum of the muon-induced events with no tagged muon and the spectrum of muon-induced events with a tagged muon, gives a measure to the percentage of the total muon-induced spectrum which is caused by untagged muons. The inefficiency in the muon veto must be proportional to this ratio. It is therefore fitted with a polynomial function below 15 keV$_{ee}$, where the inefficiency starts to become noticeable. The resulting fit is then normalised such that its value at 15 keV$_{ee}$ corresponds to 0.01. The values of the fit below 15 keV$_{ee}$ (FV) are than taken to be the inverse muon veto efficiency ($\epsilon_\mu$): 

\begin{center}
    \begin{equation}\label{eq:mu_ineff}
        \epsilon_\mu(E) = 1 - \mathrm{FV}(E)_{\mathrm{norm}}
    \end{equation}
\end{center}

\begin{figure}[bht]
    \centering
    \includegraphics[width=0.48\textwidth]{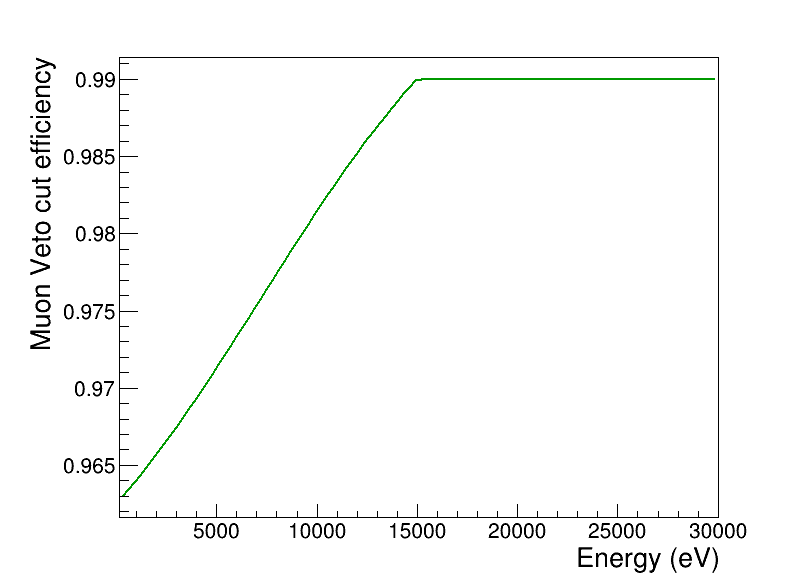}
    \caption{Final muon veto efficiency as applied to the muon simulations. For very low energies the efficiency drops to 96.5 $\%$}
    \label{fig:muon_tag_fit}
\end{figure}

In this way, it can be assured that for all energies above 15 keV$_{ee}$ an efficiency of 99.0 $\%$ is considered, while below 15 keV$_{ee}$ the tagging inefficiency is taken into account.  \newline
Figure~\ref{fig:muon_tag_fit} shows the muon veto cut efficiency that is obtained in this way. For very low energies below 400 eV$_{ee}$, it drops to 96.5 $\%$. This value is still acceptable for the CONUS+ experiment and was also approximately the value that was found for the muon veto efficiency in its predecessor experiment CONUS \cite{conus_bkg}. With these values an overall good agreement in the background model of all detectors can be found as will be shown in Section~\ref{sec:fullmodel}.  

\begin{figure}[bht]
    \centering
    \includegraphics[width=0.48\textwidth]{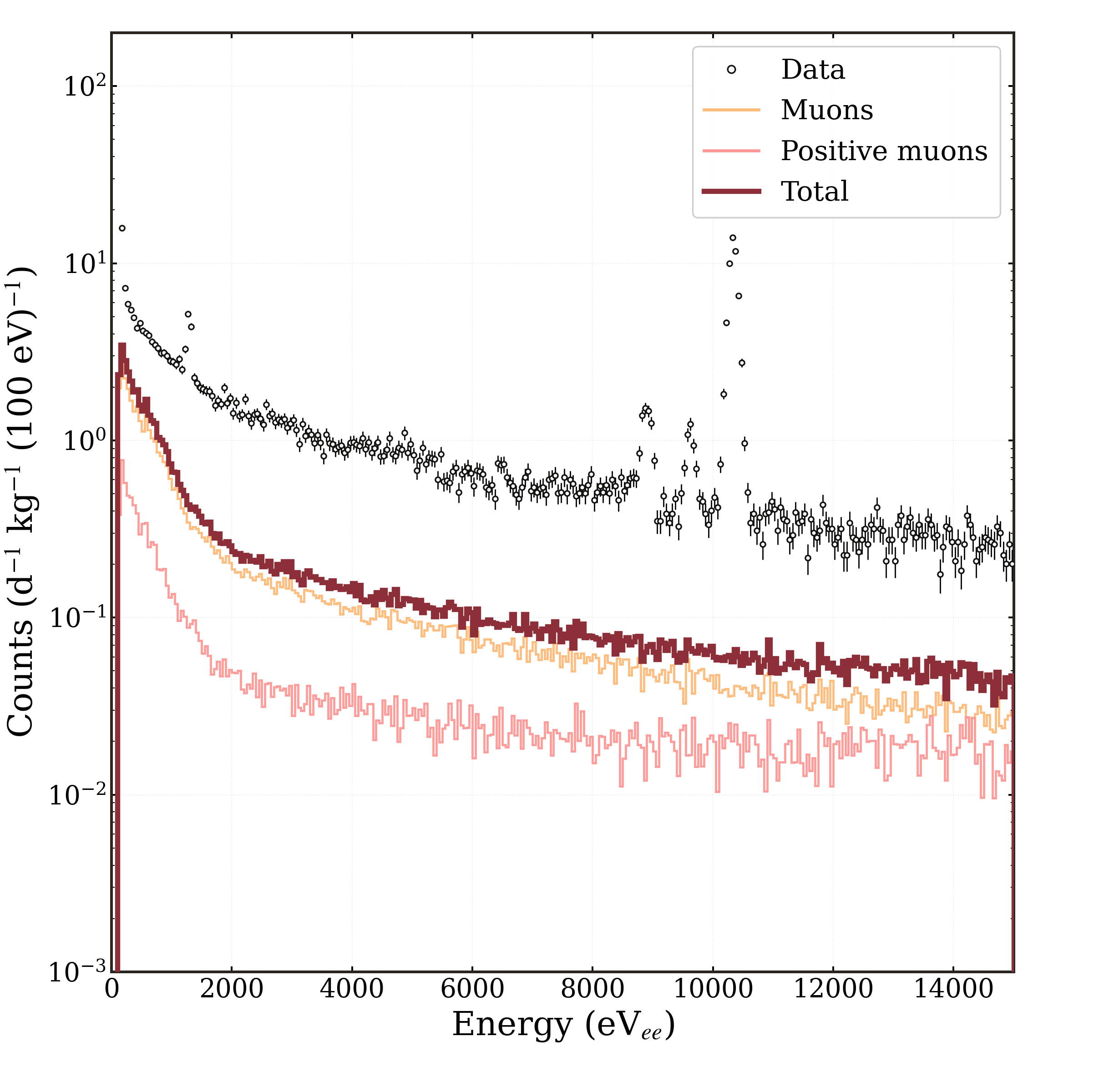}
    \caption{Muon simulation result with the applied muon veto efficiency compared to the data taken with the C5 detector in run 1. In the shown energy range the muon spectrum is flat at higher energies and rises exponentially towards low energies. This rise is due to inelastic scattering of neutrons that are induced by the muons in the high-Z materials of the CONUS+ shield.}
    \label{fig:muon_sim_with_cut}
\end{figure}

The efficiency is applied to the MC simulation results in Figure~\ref{fig:muon_comp_noVeto} and the result can be compared to the data measured in run 1 of the experiment. All detectors see the same muon flux and therefore also have the same amount of muon-induced background events in their recorded spectra, up to detector specific effects like the dead layer thickness. Figure~\ref{fig:muon_sim_with_cut} shows the comparison for the C5 detector as an example. Overall, the background contribution from muon-induced events is one of the major background sources in the experiment, similarly to what was already previously found in the CONUS experiment at KBR. Especially at low energies, a large impact of muon-induced events can be seen, an effect that mainly stems from the production of muon-induced neutrons in the lead parts of the shield.  In the energy region from 400 to 1000 eV$_{ee}$, which partially covers the CONUS+ ROI, muons contribute a background rate of (16.5 $\pm$ 1.0) counts d$^{-1}$ kg$^{-1}$ in the C2 and C3 detectors and (15.2 $\pm$ 1.0) counts d$^{-1}$ kg$^{-1}$ in the C5 detector. This corresponds to approximately 30 to 40 $\%$ of the count rate in this energy region, depending on the detector. A detailed table with the contributions in all energy regions can be found in Section~\ref{sec:fullmodel}. 

\subsubsection{High E efficiency correction with muon simulations}
\label{highEcorrection}

\begin{figure}[ht]
    \centering
    \includegraphics[width=0.47\textwidth]{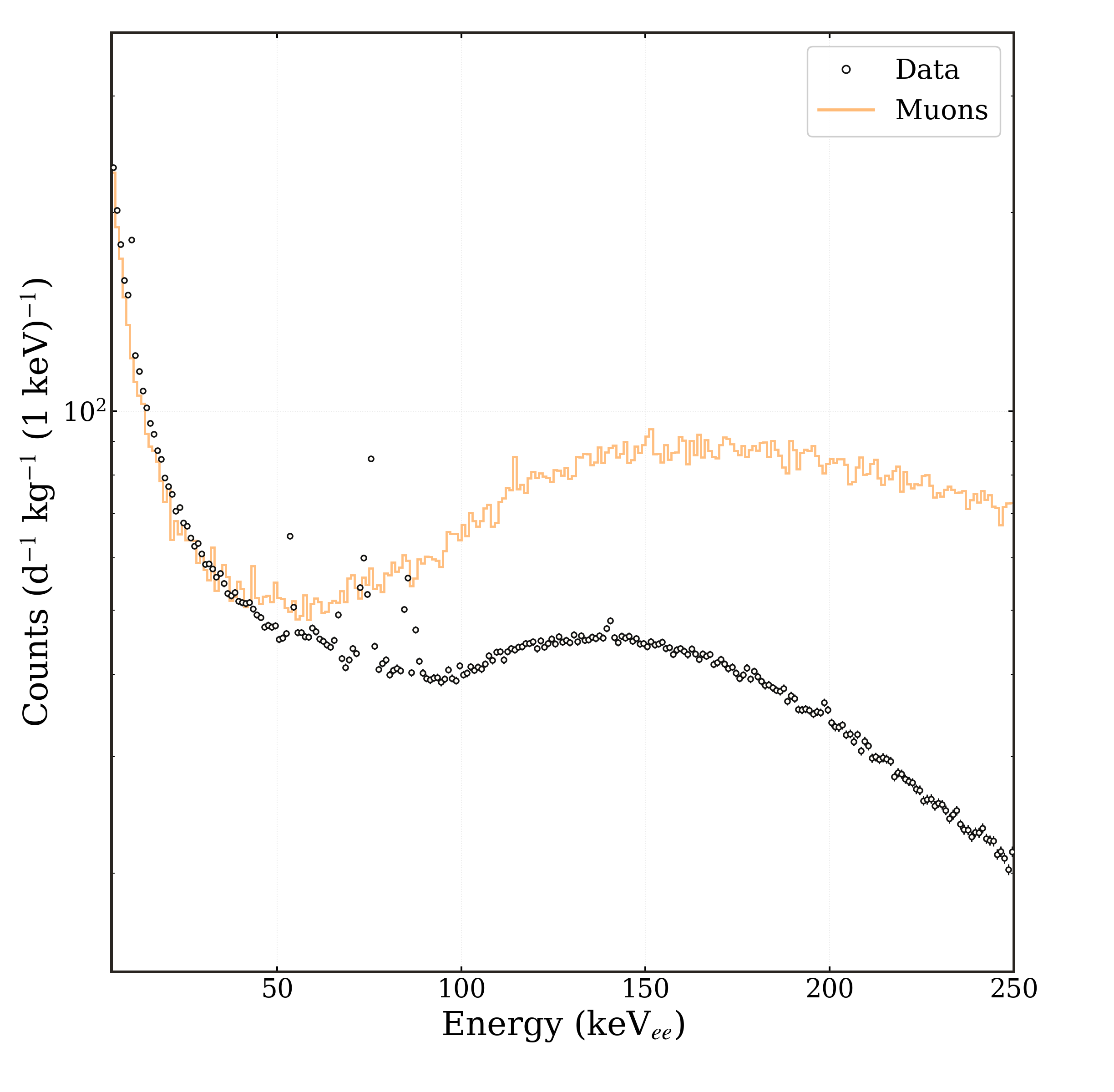}
    \caption{High E data measured in the C5 detector without muon veto cut compared to the result of the muon simulation presented in the previous sections. The data without muon veto cut should be completely dominated by muon-induced events and the simulations should therefore describe the data.}
    \label{fig:muon_highE_datavssim}
\end{figure}

As mentioned in Section~\ref{sec:data_collection}, the high energy channels of the four CONUS+ detectors show large efficiency losses in run 1. This is apparent when comparing the measured data in these channels without application of the muon veto to the muon simulation result from the previous section as shown in Figure~\ref{fig:muon_highE_datavssim}. The comparison shows good agreement up to energies of approximately 50 keV$_{ee}$, but significant deviations above which increase towards higher energies. 

\begin{figure}[ht]
    \centering
    \includegraphics[width=0.47\textwidth]{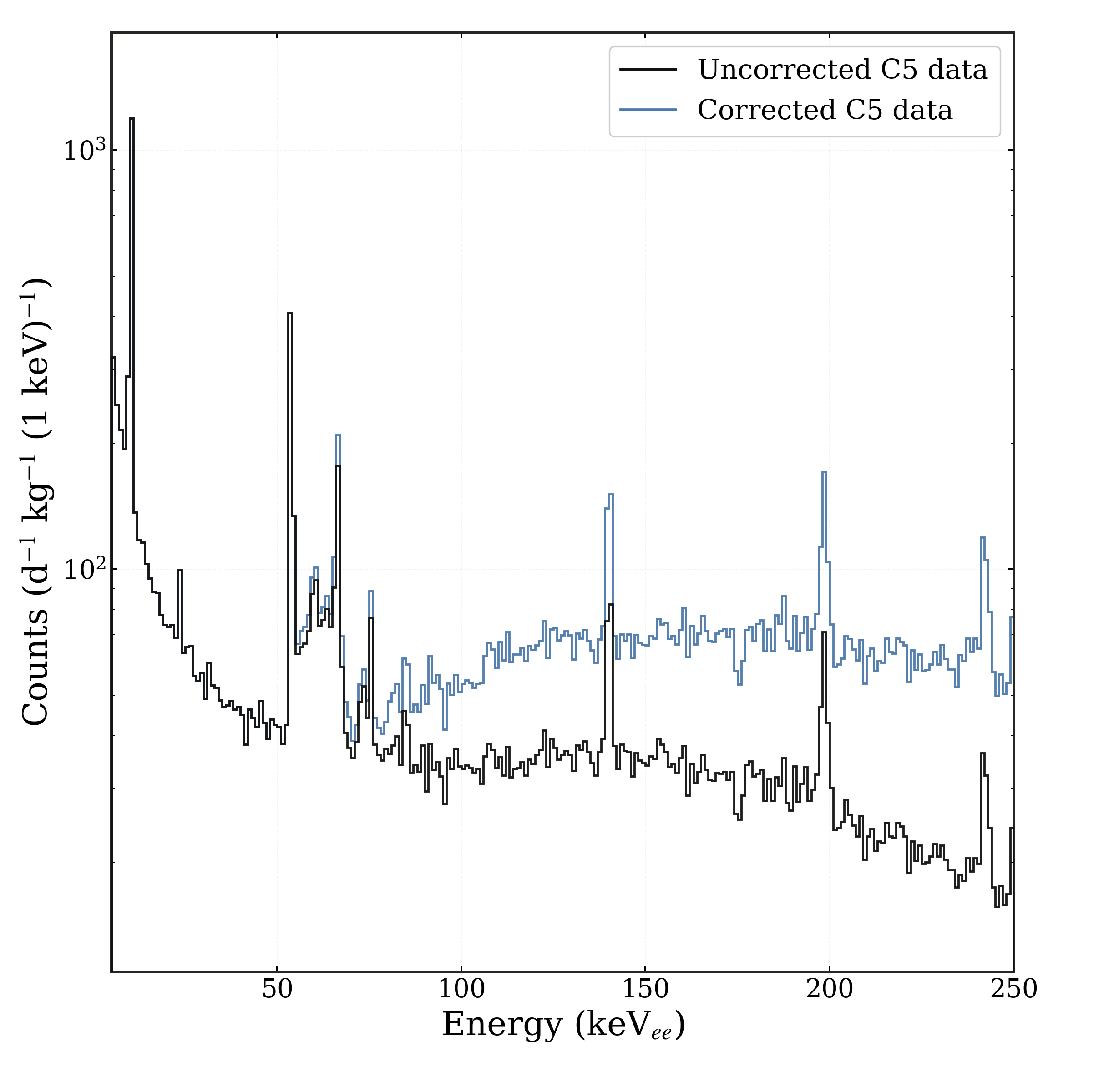}
    \caption{Corrected high energy spectrum of the C5 detector used in the analysis of run 1. The spectrum matches the expected shape after the efficiency correction with the muon simulation results. The method introduces large uncertainties in the bins above 300 keV$_{ee}$ due to the limited statistics in the original data. Therefore, only the spectra up to 300 keV$_{ee}$ are used for analysis.}
    \label{fig:muon_highE_corrected}
\end{figure}

For run 1, the efficiency loss is corrected using the muon simulation directly. It is known that the data taken in the high energy channels without application of the muon veto cut should follow the muon simulations. Therefore, the result of the simulations is divided by
the data bin-by-bin for each high energy channel. In this way, a correction factor can be assigned to each bin of the data, and the resulting correction function can be applied also to data with muon veto cut to mitigate the loss of efficiency. This method, while giving good results and producing workable spectra, is not an exact reproduction of the true high energy spectra, due to statistical fluctuations in the data (and the model) and the presence of gamma-ray lines in the measured data. To decrease the impact of this effect, the obtained correction function is smoothed with a sufficiently large bin number. The result of the correction is shown in Figure~\ref{fig:muon_highE_corrected}. The spectra of all three detectors after the correction follow the expected shapes found in the studies in \cite{conus_bkg}. The goodness of the method can also be verified by using specific gamma-ray peaks from radon decays, as will be done in Section~\ref{sec:radon}.

\subsection{Cosmic neutrons}
\label{sec:cosmicn}

\begin{figure}
    \centering
    \includegraphics[width=0.47\textwidth]{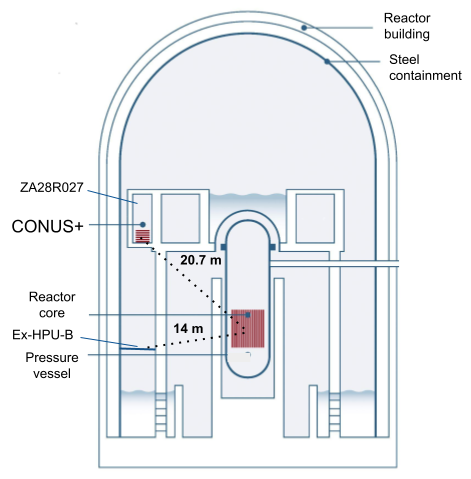}
    \caption{Schematic drawing of the reactor building at KKL with the location of the CONUS+ experiment indicated.}
    \label{fig:power_plant}
\end{figure}

The flux and spectrum of the cosmic neutrons in the CONUS+ room is determined from MC simulations of the propagation of neutrons through the reactor building. The method is well established and has been previously used in the context of other CE$\nu$NS experiments, like \cite{NUCLEUS:2017htt} and \cite{Augier_2023}. For this, in a first step a model of the reactor building was implemented in MaGe. Figure~\ref{fig:power_plant} shows a schematic drawing of the building. The important features for accurate implementation are the materials above and around the CONUS+ room and their densities to account for the correct amount of overburden. The concrete of the walls and ceiling of the room is a standard concrete with a density of 2.5 $\frac{g}{cm^3}$, while the steel of the steel containment structure is assumed to have a density of 7.85 $\frac{g}{cm^3}$. The outer wall of the reactor building is made of reinforced concrete, a standard procedure in nuclear power plants to provide earthquake safety and stability against impacts. From the technical drawing provided by KKL, it can be concluded that this reinforced concrete is made from standard concrete with additional steel elements. The exact density profile is not known but can be calculated by requiring the complete structure above the CONUS+ room to have an effective overburden of the measured (7.3 $\pm$ 0.1) m w.e. In this way, a density of (5.2 $\pm$ 0.1) $\frac{g}{cm^3}$ can be estimated.

\begin{figure}[ht]
    \centering
    \includegraphics[width=0.47\textwidth]{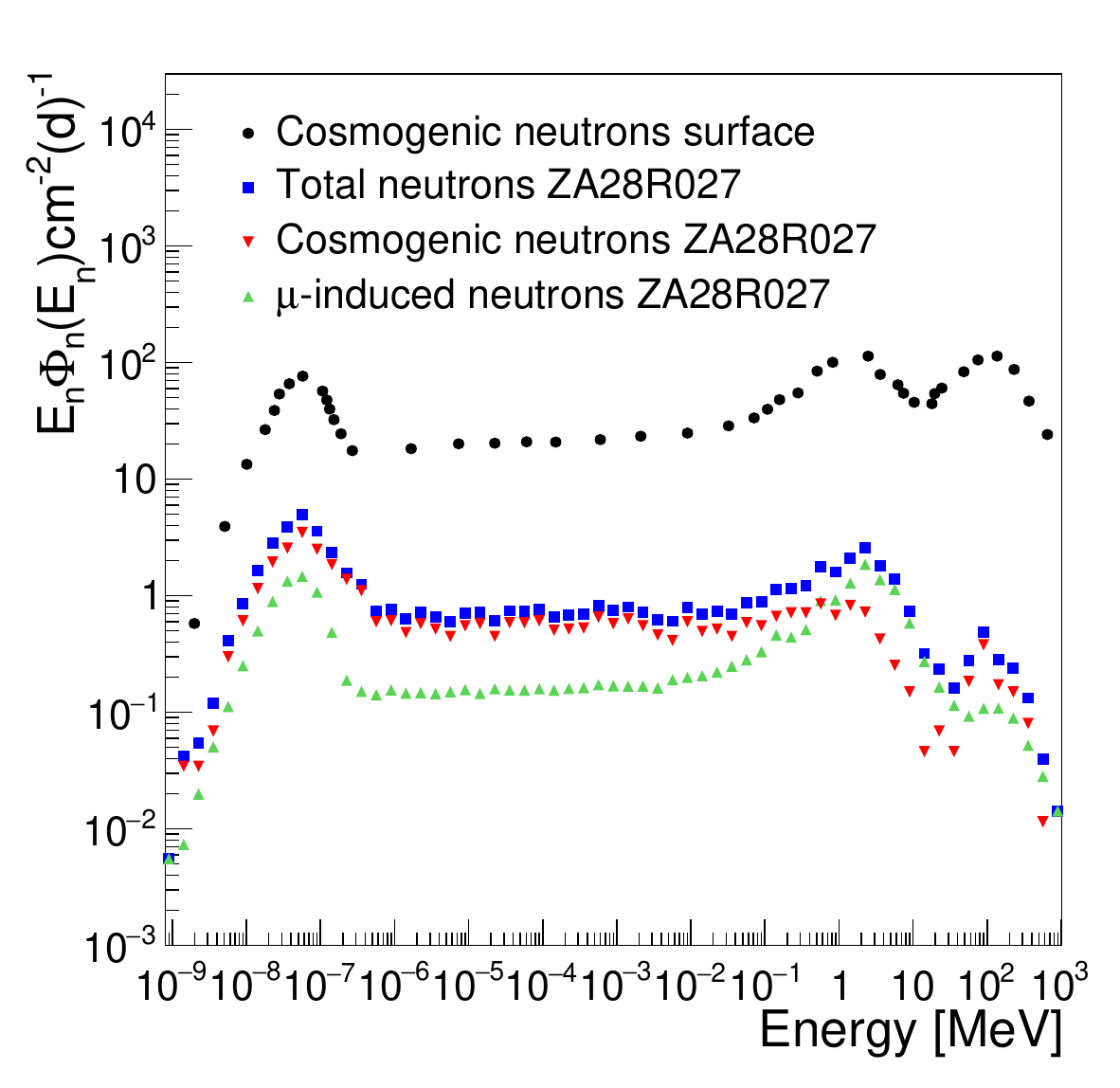}
    \caption{Neutron flux in the CONUS+ room originating from cosmic rays, as shown in \cite{conusplus_bkg}. Total simulated neutron spectra are shown in blue. The muon-induced neutrons generated in the reactor building (green) produce a peak around 1~MeV. The cosmogenic neutrons (red) are reduced by almost 2 orders of magnitude with respect to the surface (black). A peak around 100~MeV is produced by the latter component.}
    \label{fig:neutron_flux}
\end{figure}

In addition, the cosmic neutron flux at the location of KKL in Switzerland is needed. Unfortunately, such a measurement is not available in the literature and instead a neutron flux measurement from a comparable location is used \cite{Gordon:2004}. In their work, Gordon et al. performed several neutron measurements at different locations in the continental USA. One of these measurements was performed at Yorktown Heights, NY at an altitude of 170 m over sea level. The spectrum is shown in black in Figure~\ref{fig:neutron_flux}. The flux measured in the Yorktown measurement was 0.0134 cm$^{-2}$ s$^{-1}$. Due to the difference in altitude between Yorktown and Leibstadt (350 m over sea level) a small correction is applied to the measured flux and this work uses a value of (0.0142 $\pm$ 0.0020)~cm$^{-2}$~s$^{-1}$ in the following MC simulations. The correction also takes into account the different locations of the measurements in terms of geographical latitude and longitude. No change in the principal shape of the spectrum is expected \cite{Gordon:2004}. The spectrum shows three characteristic peaks, which are typical of cosmic neutron measurements: a thermal peak at around 10$^{-8}$ MeV, where neutrons have been slowed down to the point where they are in equilibrium with the nuclei of the atmosphere, an evaporation peak (E $\sim$ 1 MeV), and the cosmic/fast neutron peak (E $>$ 10 MeV). The cosmic neutron peak is of particular interest to the background of the CONUS+ experiment since these neutrons can in principle penetrate the shield.

For the MC simulations, 10$^{11}$ neutrons with the blue energy distribution in Figure~\ref{fig:neutron_flux} and isotropic angular distribution are started from a 50 m x 50 m square plane above the reactor building. The neutrons and their energies are registered in a half-sphere placed at the location of the CONUS+ shield in the building model. The resulting energy distribution is shown in red in Figure~\ref{fig:neutron_flux}.

\begin{figure}[htbp]
  \centering
  \begin{subfigure}[b]{0.47\textwidth}
    \centering
    \includegraphics[width=\textwidth]{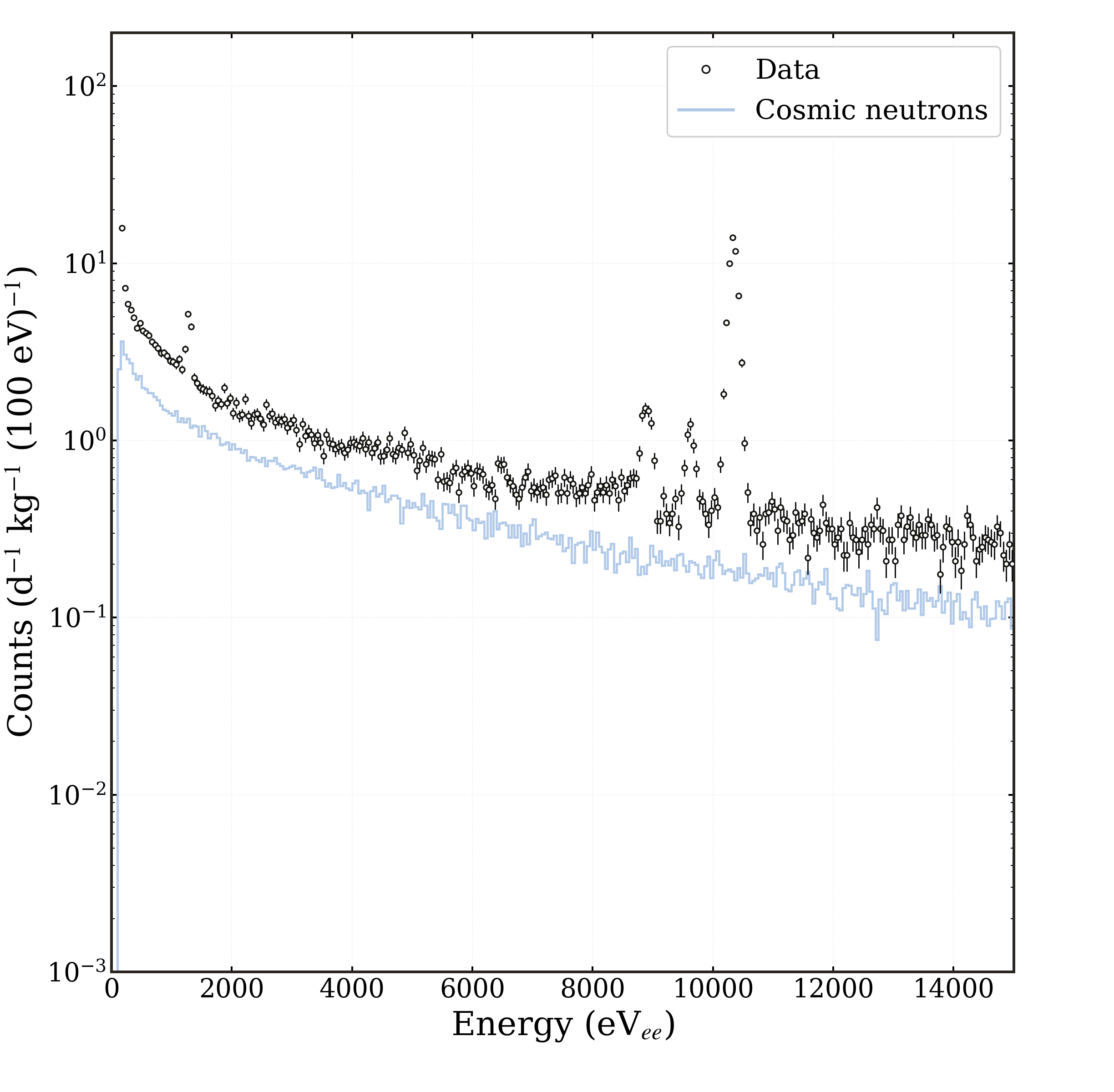}
    \label{fig:neutrons_sim_result_lowE}
  \end{subfigure}
  \hfill
  \begin{subfigure}[b]{0.47\textwidth}
    \centering
    \includegraphics[width=\textwidth]{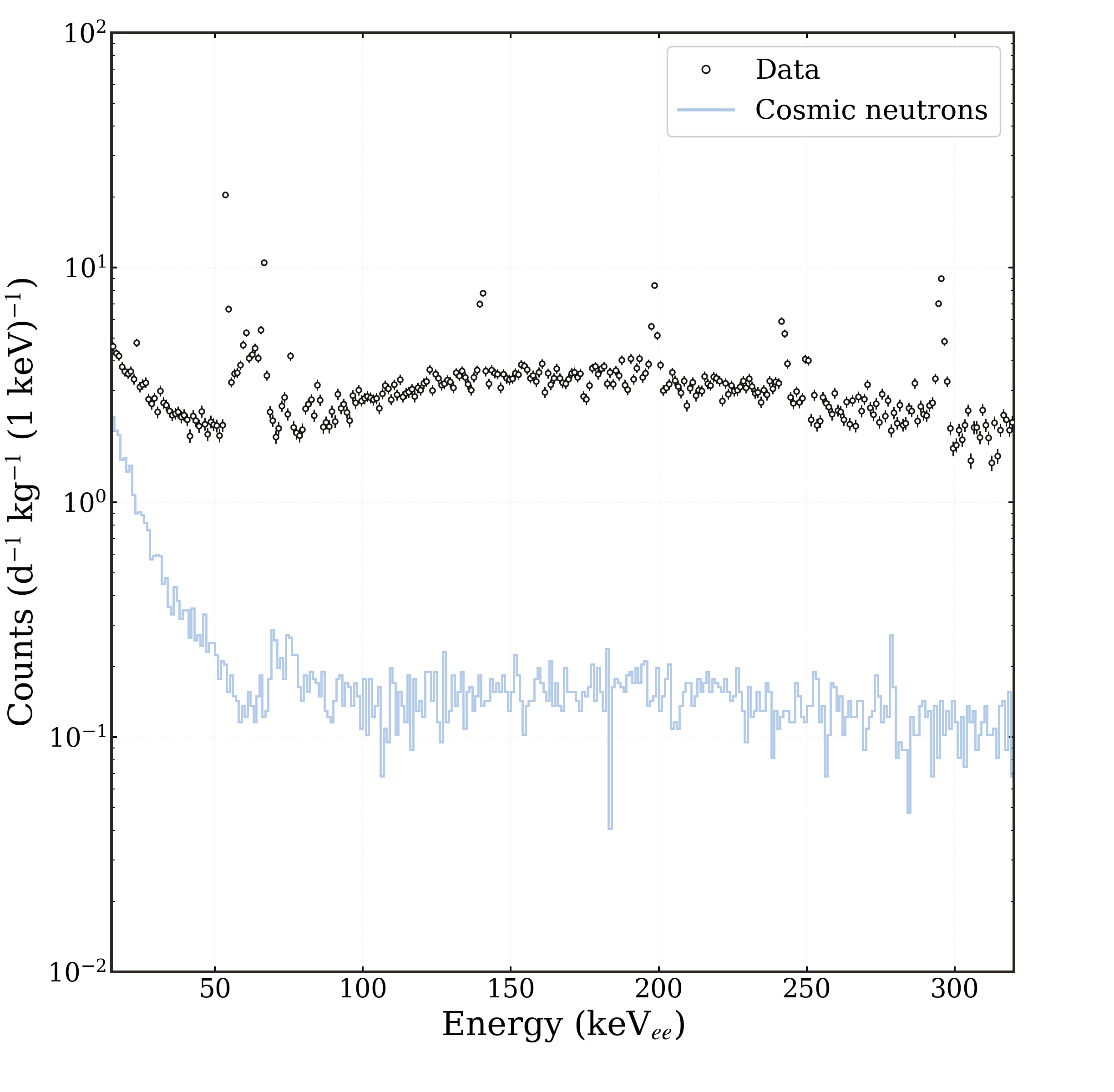}
    \label{fig:neutrons_sim_result_highE}
  \end{subfigure}
    \caption{Neutron simulation result compared to the data taken in run 1 of the CONUS+ experiment with the C5 detector in the low energy channel up to 30 keV$_{ee}$ (top) and the high E channel (bottom). The neutrons feature an exponential rise towards very low energies and their spectrum flattens towards higher energies.}
  \label{fig:neutrons_sim_result}
\end{figure}

The total flux of neutrons that arrive in the CONUS+ room is found to be (0.9 $\pm$ 0.2)~cm$^{-2}$~d$^{-1}$ with the uncertainty being propagated from the uncertainty of the assumed outside flux at the KKL location, as well as from uncertainties in the exact composition of the reinforced concrete in the KKL reactor building. The latter were assessed by performing several MC simulations with varying densities of the overburden materials. The shape of the spectrum is consistent with other simulated neutron spectra in similar shallow depth experiments, like in \cite{Augier_2023}. As expected, the flux is greatly reduced compared to the outside flux by around two orders of magnitude, however, a small flux remains. This is especially critical, since a fraction of very fast neutrons (E $>$ 10 MeV) originating from the fast neutron peak remain in the CONUS+. The reduction from the overburden is large enough to avoid activation of certain radioactive isotopes like $^{68}$Ge (see Section~\ref{sec:activation}) in the germanium detectors especially considering the additional 20 cm of lead of the CONUS+ shield. \newline
In order to study the effect of the remaining neutrons on the background, a final MC simulation is performed in which neutrons with the red energy distribution in Figure~\ref{fig:neutron_flux} and an isotropic angular distribution are started from the walls of the CONUS+ room. Figure~\ref{fig:neutrons_sim_result} shows the results of this MC simulation compared to the low and high energy channels of the C5 data in run 1.

The MC simulations show an exponential rise of the background contribution from cosmic neutrons towards low energies coming from elastic scattering of the neutrons on the germanium nuclei in the detectors. As such, they make up to 50 \% of the background rate in the energy region from 400 to 1000 eV$_{ee}$, as shown in Table~\ref{tab:neutron_bkg_contributions_on}. Cosmic neutrons are therefore one of the dominant sources of background in the ROI of CONUS+.  

\begin{table}[bht]
\caption{Contribution to the background count rates in the three detectors from neutrons in the energy region from 400 to 1000\,eV$_{ee}$.}
\label{tab:neutron_bkg_contributions_on}
\centering
\small
\setlength{\tabcolsep}{5pt}
\setlength{\extrarowheight}{2pt}

\begin{tabular}{lcc}
\hline
Detector &
\begin{tabular}[c]{@{}c@{}}Count rate\\ {[}400,\,1000{]}\,eV$_{ee}$\\ (d$^{-1}$\,kg$^{-1}$)\end{tabular} &
\begin{tabular}[c]{@{}c@{}}Fraction of total\\ background (\%)\end{tabular} \\
\hline
C5 & 21.6 $\pm$ 3.1 & 50.3 $\pm$ 7.2 \\
C2 & 21.6 $\pm$ 3.1 & 41.3 $\pm$ 5.9 \\
C3 & 21.6 $\pm$ 3.1 & 43.8 $\pm$ 6.2 \\
\hline
\end{tabular}
\end{table}

\subsection{Muon-induced neutrons in the overburden}
\label{sec:muoninducedn}
The last component originating from interactions of cosmic rays to consider for the background model are muon-induced neutrons produced from photonuclear interactions, muon capture, or deep-inelastic scattering in the overburden of the experiment. These neutrons were not considered in the previous two sections. To investigate them, muons were propagated through the model of the reactor building in an analogous way to the procedure described for the cosmic neutrons in the previous section. The initial muon spectrum was again calculated from \cite{reyna, Bugaev}. The resulting neutron flux in the CONUS+ room is shown in green in Figure~\ref{fig:neutron_flux}. The total flux of muon-induced neutrons in the CONUS+ room is around one order of magnitude smaller than that of the cosmic neutrons and noticeably does only include a small fraction of muons in the fast neutron peak at energies above 10 MeV. 

The flux is again propagated through the shield in the same procedure as for the cosmic neutrons. Due to the relative reduction of the fast neutron peak and also the overall smaller neutron flux, the impact of these neutrons on the CONUS+ background is considerably  smaller than that of the cosmic neutrons. They contribute (2.2 $\pm$ 0.1) counts d$^{-1}$ kg$^{-1}$ in the energy region between 400 and 1000 eV$_{ee}$ (5 \% of the overall count rate) and the contribution drops off sharply at higher energies.

\subsection{Muon-induced metastable germanium states}
\label{sec:meta}

\begin{figure}[ht]
    \centering
    \includegraphics[width=0.47\textwidth]{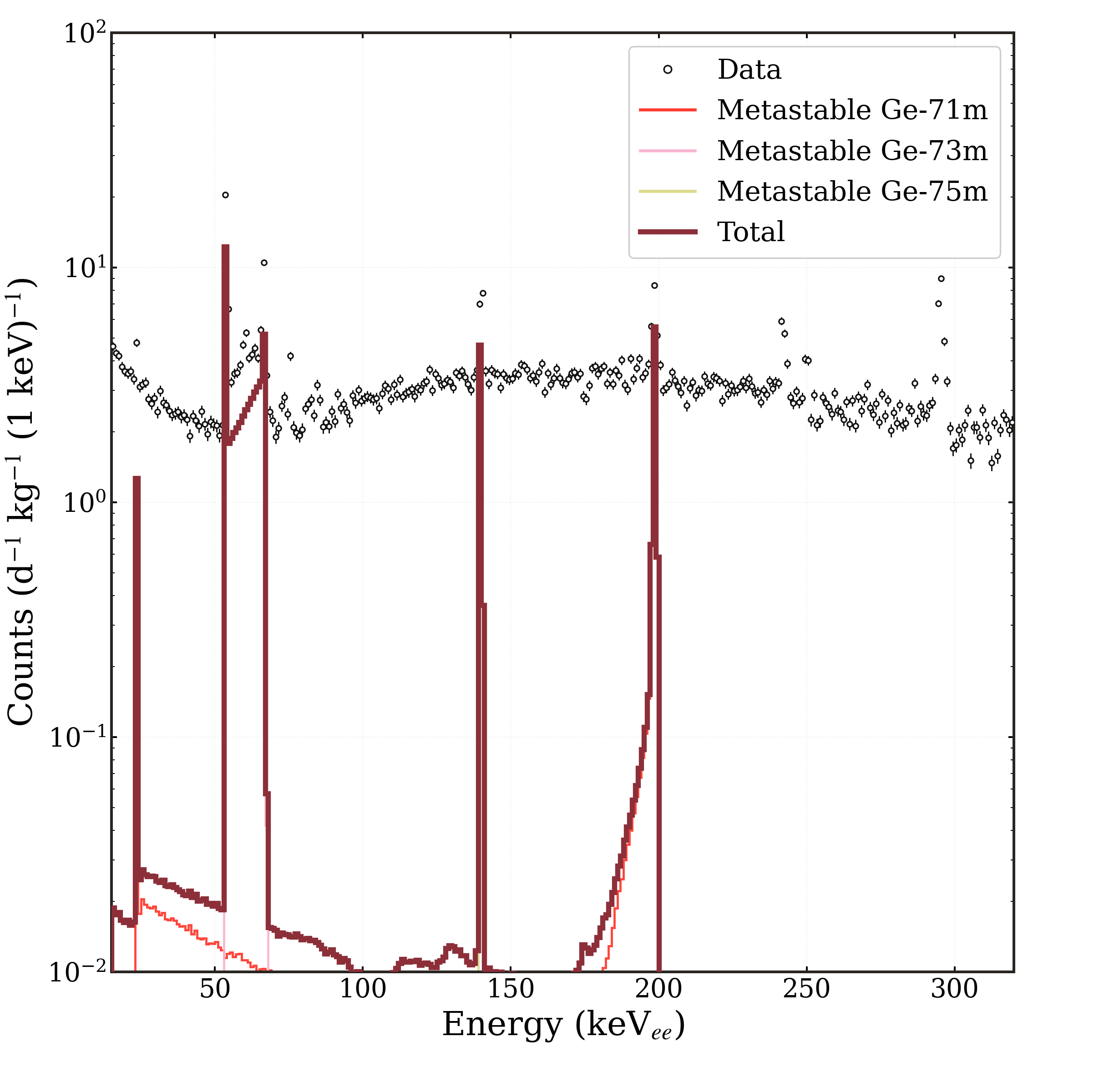}
    \caption{MC simulation result of metastable germanium states in the C5 detector. }
    \label{fig:metastable_result}
\end{figure}

As seen in Figure \ref{fig:muon_highE_corrected}, the high energy spectrum of the CONUS+ detectors show several lines. While some of these are induced by radon daughters in the detector chamber of the shield (see Section \ref{sec:radon}), the gamma-ray peaks at 140 keV, 175 keV, 198 keV,  as well as the double peak structure at 53 keV, stem from the decay of metastable germanium isotopes in the detectors. These isotopes, namely $^{71m}$Ge, $^{73m}$Ge, and $^{75m}$Ge,  are produced from neutron capture of muon-induced neutrons from the shield and they feature live-times larger than the muon veto window. They are not directly produced in the MC muon simulations in MaGe and also are too long-lived to be cut by the muon veto system, which is why a dedicated simulation was performed. The results are found in Figure \ref{fig:metastable_result}.

The double peak structure, seen from the decay of $^{73m}$Ge, originates from the very short half-life of only 2.9 $\mu$s between the emission of the 53.3 keV line and the 13.3 keV line in the decay of this state, which in many cases can not be resolved in the detectors. If the emission of the 13.3 keV gamma happens within the decay of the pulse created by the initial 53.3 keV gamma, they are registered as one event in the DAQ system of the experiment, with a combined energy depending on the exact timing of the second gamma emission. More details can be found in \cite{Ackermann2025CONUS}. The overall background contribution from the metastable germanium states is very small in all energy ranges where they do not feature their respective gamma-ray peaks. In the energy range from 0.4 to 1 keV$_{ee}$, they contribute below 0.1 $\%$ of the total background rate for all detectors.

\section{Radon in the detector chamber}
\label{sec:radon}

The inside of the CONUS+ shield, the detector chamber, is continuously being flushed with radon-free air. However, as can be seen, for example, from Figure~\ref{fig:neutrons_sim_result}, the measured high energy spectra of all three detectors still show three gamma-ray peaks from $^{214}$Pb, a daughter nucleus of radon, at 242, 295, and 352 keV. The remaining radon impact on the background is therefore modelled in the following. \newline 
In a first step the count rates of the three lines in the data is calculated by performing a Gaussian fit to the lines and subtracting the continuous background. The resulting count rates are listed in Table~\ref{tab:radon_lines_meas}. The differences in the count rates between detectors stem from the fact that the flushing with radon-free air introduces inhomogeneities in the distribution of radon in the detector chamber. 

\begin{table*}[t]
\caption{Count rates of the radon induced lines in the high energy channels of the three detectors used in the run 1 analysis. The values for the count rates in the 352 keV line (marked with *) are not completely reliable, since at these high energies the efficiency correction with the muon simulation fails due to the low number of statistics in the original data, as explained in Section~\ref{highEcorrection}. The uncertainties given for the count rates only considers the poissonian uncertainty from the extracted value. }
\label{tab:radon_lines_meas}
\centering
\setlength\extrarowheight{4pt}
\begin{tabular}{|l| ccc|}
\hline
Detector & C5  & C2  & C3  \\
\hline
Energy in data [keV$_{ee}$] & 242.085 $\pm$ 0.024 &  241.967 $\pm$ 0.029 & 242.158 $\pm$ 0.020\\
Count rate [d$^{-1}$ kg$^{-1}$]& 5.77 $\pm$ 0.14 &  8.49 $\pm$ 0.29 & 7.64 $\pm$ 0.07\\
\hline
Energy in data [keV$_{ee}$] & 295.229 $\pm$ 0.012 &  295.275 $\pm$ 0.016 & 295.468 $\pm$ 0.008\\
Count rate [d$^{-1}$ kg$^{-1}$]& 16.63 $\pm$ 0.31 &  20.70 $\pm$ 0.40 & 19.8 $\pm$ 0.29\\
\hline
Energy in data [keV$_{ee}$] & 352.033 $\pm$ 0.006 &  351.922 $\pm$ 0.011 & 352.177 $\pm$ 0.001\\
Count rate [d$^{-1}$ kg$^{-1}$]& 24.61 $\pm$ 0.45 (*) &  26.77 $\pm$ 0.47 (*)& 20.08 $\pm$ 0.38 (*)\\
\hline
\end{tabular}
\end{table*}

The relative intensity of the three radon-induced lines can also be used to cross check the muon-simulation-based efficiency correction in the high E channels of the detectors (see Section~\ref{highEcorrection}). Table~\ref{tab:radon_lines_ratio} shows a comparison between the ratio of the intensities of the lines and values found in literature. For the ratio between the 242 and 295 keV line good agreement is found, while the ratio between the 352 and 295 keV line is consistently too low in all detectors which is due to the unreliability of the efficiency correction for energies above 300 keV$_{ee}$.

\begin{table}[bht]
\caption{$\gamma$-ray lines from radon decay chain in the CONUS+ high energy channels. Literature values are taken from \cite{nudat3}}
\label{tab:radon_lines_ratio}
\centering
\setlength\extrarowheight{4pt}
\begin{tabular}{lcc}
\hline
& \shortstack{Ratio 295 keV line \\ to 242 keV line} 
& \shortstack{Ratio 352 keV line \\ to 295 keV line} \\
\hline
Literature  & 2.544 $\pm$ 0.021 &  1.934 $\pm$ 0.017\\
C5 data & 2.88 $\pm$ 0.09 &  1.48 $\pm$ 0.04\\
C2 data & 2.44 $\pm$ 0.10 &  1.29 $\pm$ 0.03\\
C3 data & 2.59 $\pm$ 0.09 &  1.01 $\pm$ 0.02\\
\hline
\end{tabular}
\end{table}

\begin{figure}[htbp]
  \centering
  \begin{subfigure}[b]{0.47\textwidth}
    \centering
    \includegraphics[width=\textwidth]{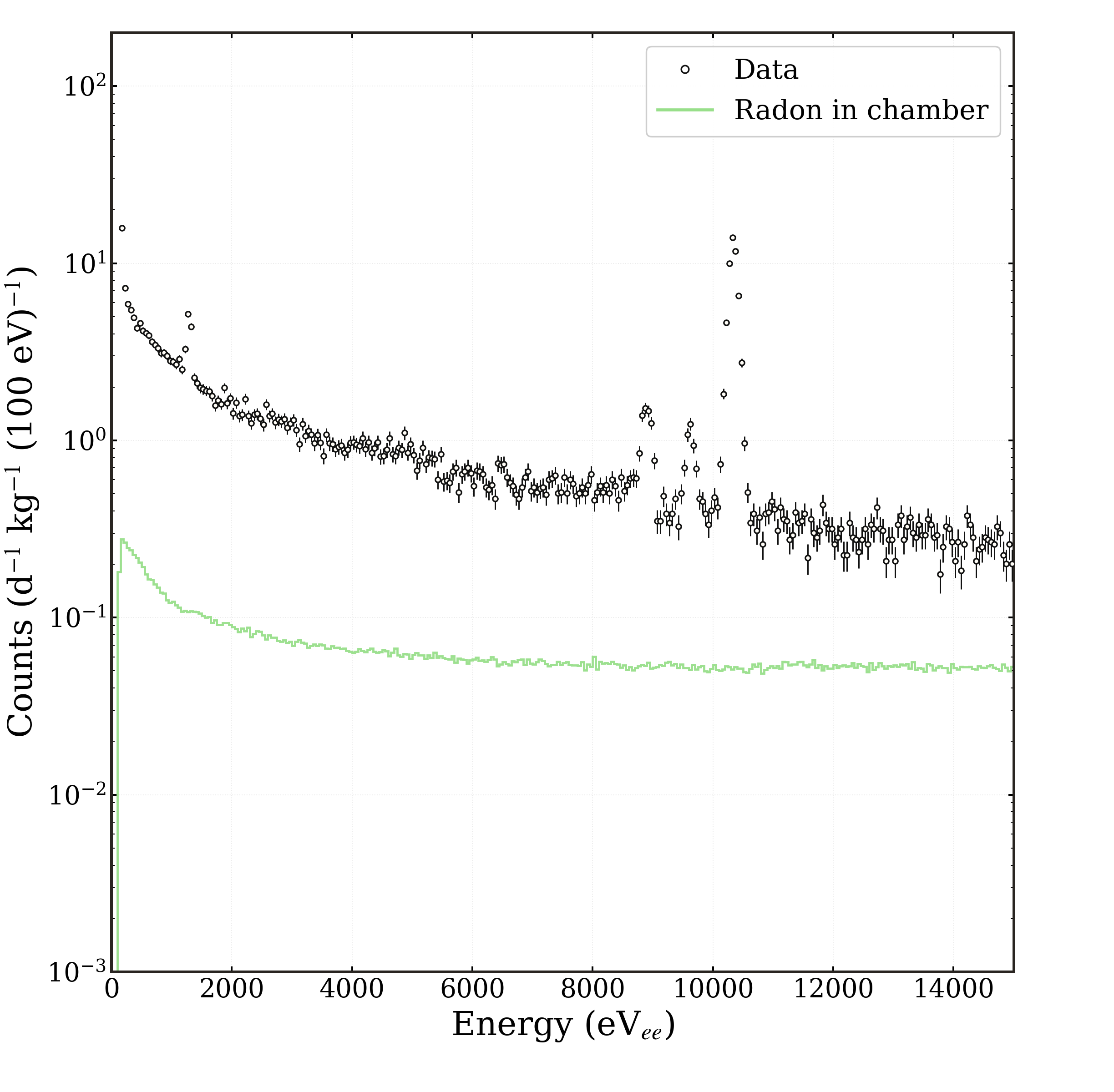}
    \label{fig:radon_sim_result_lowE}
  \end{subfigure}
  \hfill
  \begin{subfigure}[b]{0.47\textwidth}
    \centering
    \includegraphics[width=\textwidth]{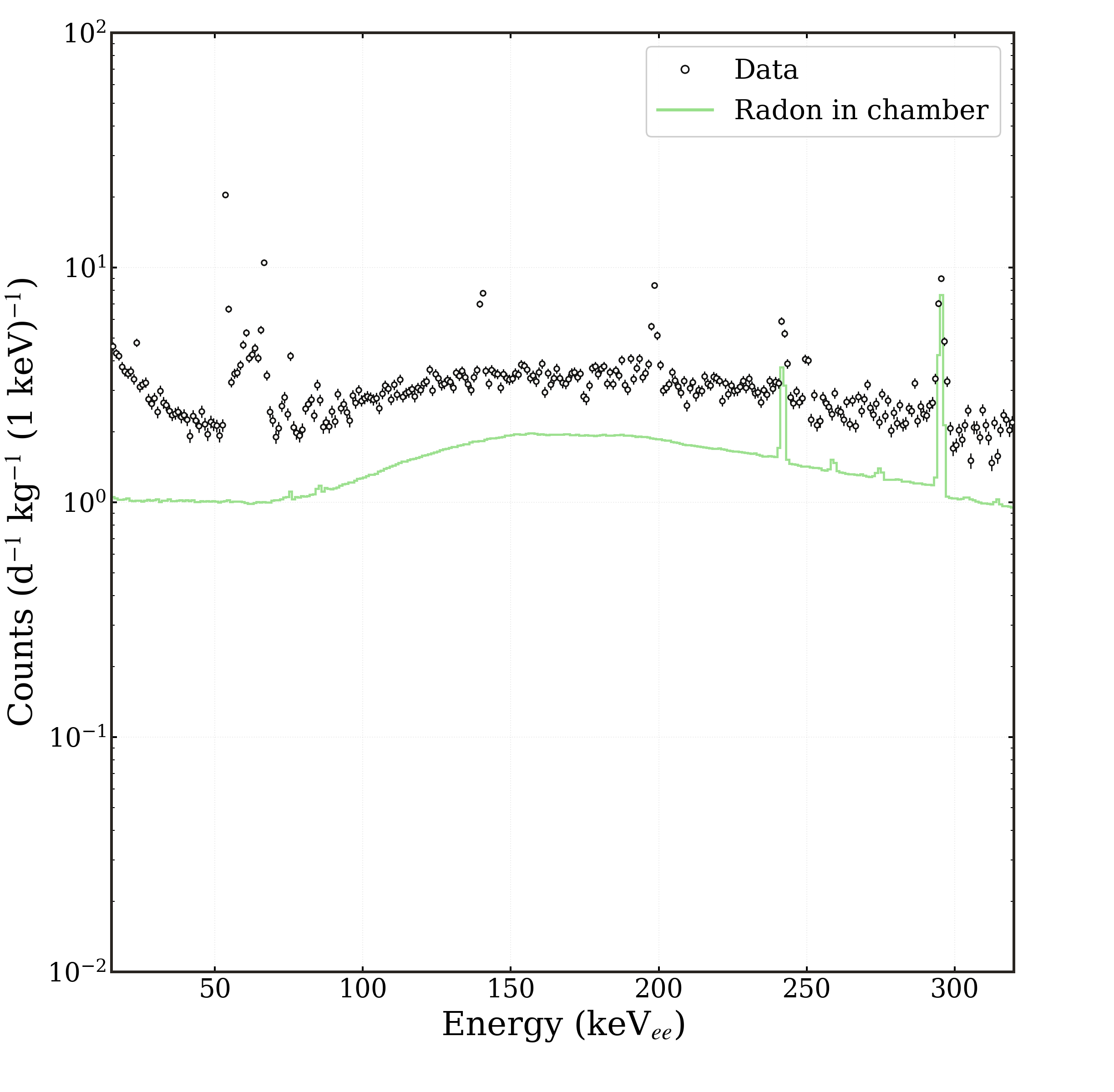}
    \label{fig:radon_sim_result_highE}
  \end{subfigure}
    \caption{MC simulation result for radon in the detector chamber compared to the data taken in run 1 of the CONUS+ experiment with the C5 detector in the low energy channel up to 30 keV$_{ee}$ (top) and the high E channel (bottom). The impact of radon is small in the low energy channel and the ROI but is considerable and one of the largest contributions in the high energy range. }
  \label{fig:radon_sim_result}
\end{figure}

For the MC simulation of the radon component, $^{214}$Pb and $^{214}$Bi isotopes are placed on the bottom and the walls of the detector chamber and the end caps of the detector cryostats. The results of the MC simulations are then scaled such that the count rates in the $^{214}$Pb peaks are in agreement with the count rates in Table~\ref{tab:radon_lines_meas}. Secular equilibrium is assumed. The MC simulation result is shown for the C5 detector in Figure~\ref{fig:radon_sim_result}. The impact in the ROI of the experiment is subdominant and contributes around 5 $\%$ of the background rate in the 400 to 1000 eV$_{ee}$ range. The contribution at high energies between 50 and 300 keV$_{ee}$ is considerably higher (up to 50 $\%$).

\section{Cosmogenic activation}
\label{sec:activation}

Cosmogenic activation describes the process by which radioactive isotopes are induced in materials when these are exposed to cosmic rays. The main processes responsible for this production are neutron capture and neutron-/proton-induced spallation. Spallation is only possible for primary energies above 10 MeV, which is why activation through this interaction is only induced when materials are directly exposed to cosmic rays. The overburden of the CONUS+ experiment in combination with the shield is enough to prevent these interactions. Neutron capture on the other hand is dominant for neutrons with lower energies and thermal neutrons in particular. The isotopes created in this process can therefore still be produced in materials inside the shield, primarily from muon-induced neutrons.  For the case of the CONUS+ experiment, two materials are of particular interest: the germanium crystals and the copper parts of the cryostat. For the copper elements, the most relevant isotopes are $^{60}$Co (T$_{1/2}$ = 1923.6 d \cite{NuDat}), produced from neutron capture on $^{59}$Co, $^{57}$Co (T$_{1/2}$ = 271.8 d \cite{NuDat}), and $^{54}$Mn (T$_{1/2}$ = 312.2 d \cite{NuDat}), both produced from spallation of $^{63}$Cu and $^{65}$Cu. For the case of germanium a larger number of isotopes is relevant (see Table~\ref{tab:table_ge_cosmo}).

\begin{table}[bht]
\caption{Most relevant cosmogenically induced isotopes in germanium for CONUS+ and their production mechanism. Half lifes are taken from \cite{NuDat}.}
\label{tab:table_ge_cosmo}
\centering
\setlength\extrarowheight{4pt}
\begin{tabular}{lcc}
\hline
Isotope & Half-life [d] & Production mechanism  \\
\hline
$^{60}$Co & 1923.6 & Neutron capture on $^{59}$Co  \\
$^{57}$Co & 271.8 & Spallation of $^{63}$Cu and $^{65}$Cu   \\
$^{54}$Mn & 312.2 & "  \\
$^{71}$Ge & 11.4 & Neutron capture on $^{70}$Ge   \\
$^{68}$Ge & 271.0 & Spallation of different Ge nuclei  \\
$^{68}$Ga & 0.05 & Decay of $^{68}$Ge  \\
$^{65}$Zn & 244.0 & Neutron capture on $^{64}$Zn  \\
$^{3}$H & 4493.9 & Spallation of Ge nuclei  \\
$^{55}$Fe & 1002.7 & Spallation of $^{56}$Fe  \\
\hline
\end{tabular}
\end{table} 

\begin{figure}[ht]
    \centering
    \includegraphics[width=0.47\textwidth]{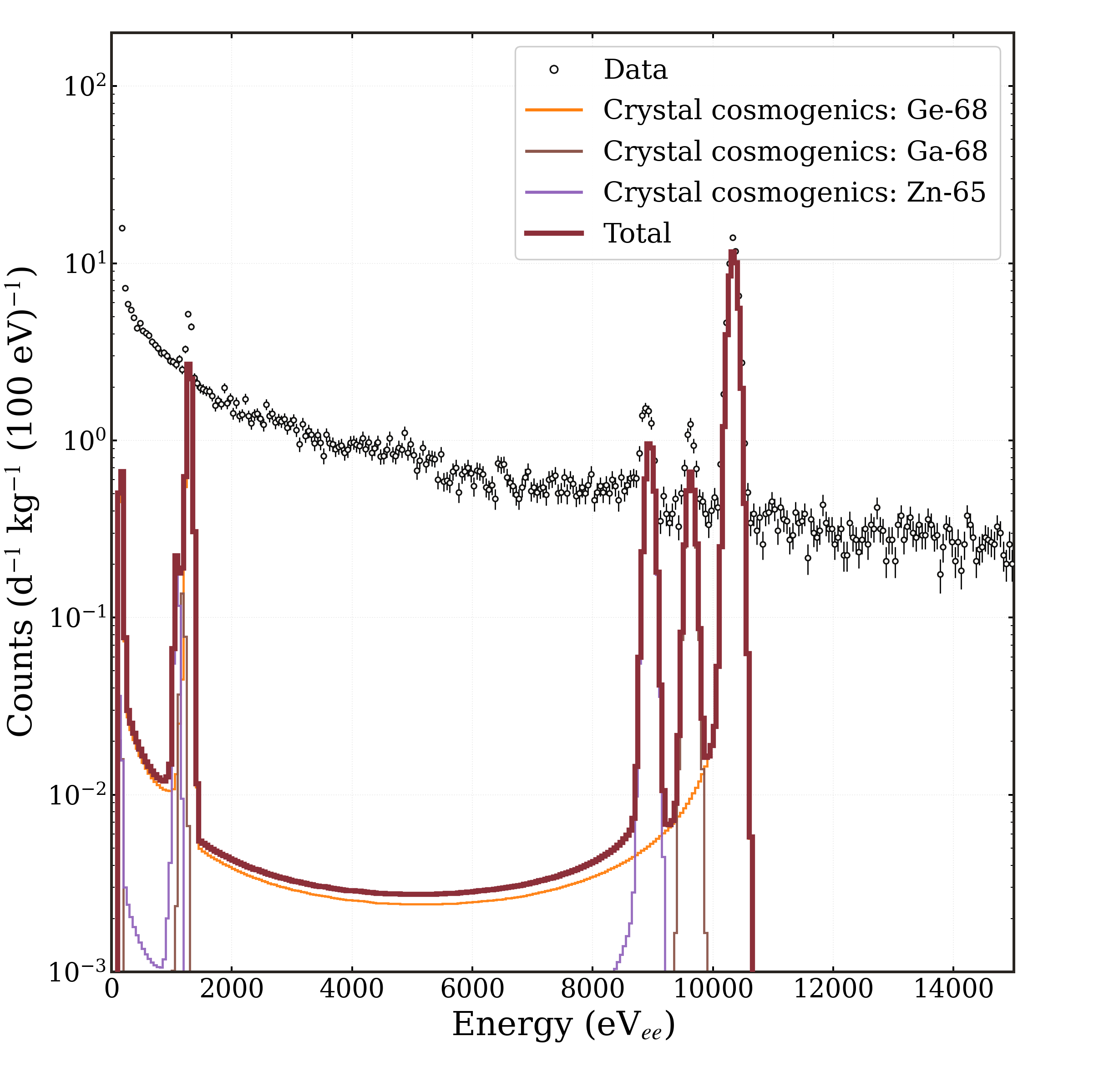}
    \caption{MC simulation results of $^{68/71}$Ge, $^{68}$Ga, and $^{65}$Zn for the C3 detector. }
    \label{fig:ge_cosm_lowE_lines}
\end{figure}

Of the listed isotopes, only four produce visible lines in the measured CONUS+ data of Run 1: $^{68/71}$Ge at 10.37 keV, $^{68}$Ga at 9.7 keV, and $^{65}$Zn at 8.9 keV. All three lines originate from X rays produced in K shell transitions in the isotopes. All three isotopes also produce L shell lines at approximately 1.3 keV, which are visible as one shared peak in the CONUS+ data. Additionally, the $^{68/71}$Ge isotopes also produce an M-shell X ray line at very close to the lowest threshold of the CONUS+ run 1 detectors at 160 eV. The line was recently measured for the first by the CONUS+ experiment in \cite{mshell} at (158.7 $\pm$ 1.7) eV. The line contributes around 7 $\%$ of the background in the lowest energy bin of the C3 detector. The MC simulation of these visible lines follows the same procedure as for the radon simulations in the previous section. The MC simulation results are shown in Figure~\ref{fig:ge_cosm_lowE_lines}.

\begin{figure}[ht]
    \centering
    \includegraphics[width=0.47\textwidth]{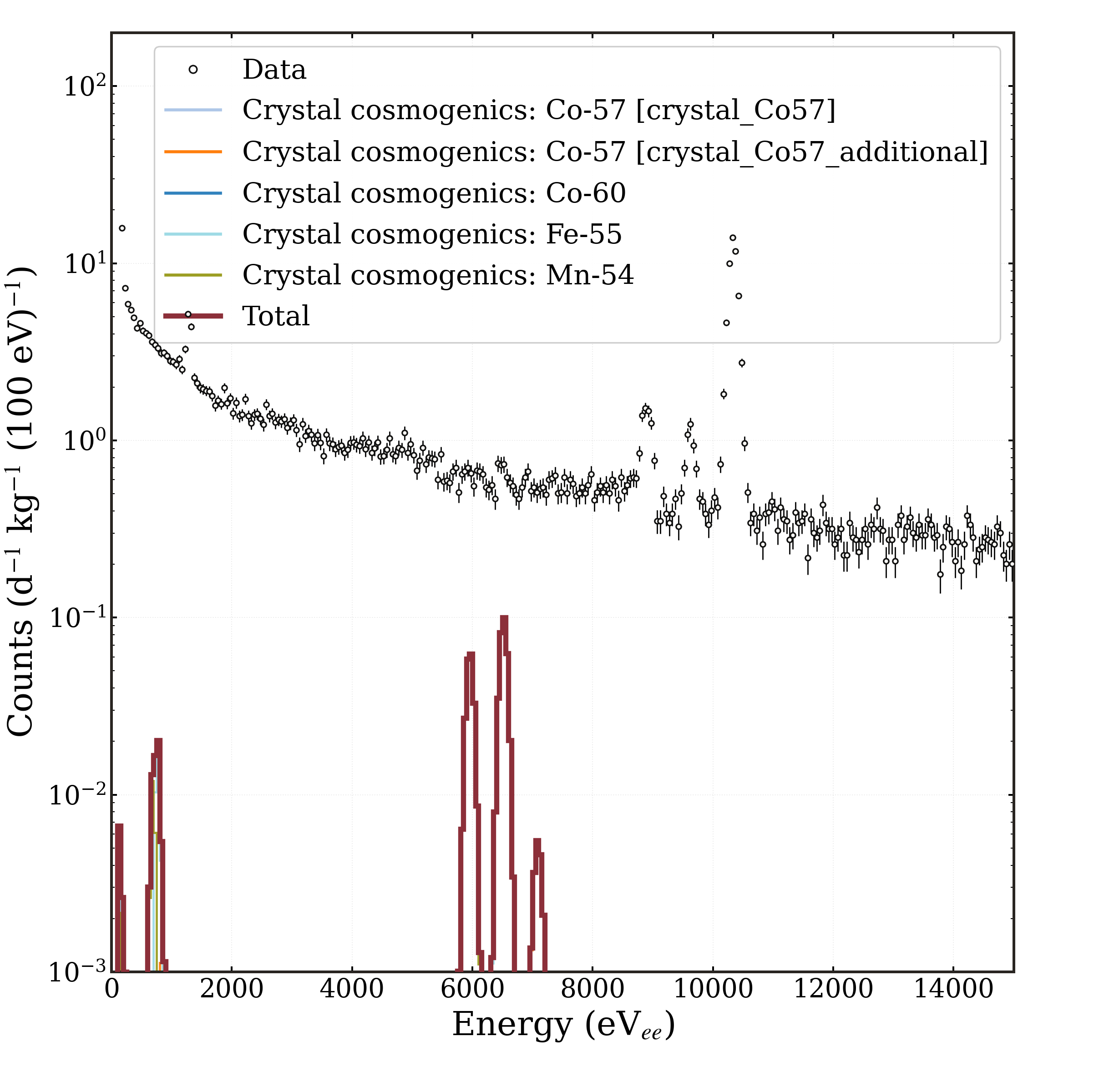}
    \caption{Background contributions from cosmogenically activated isotopes in the germanium crystals without visible lines. Their corresponding activities were calculated according to \cite{conus_bkg} assuming an exposure of 100 days. }
    \label{fig:ge_cosm_no_lines}
\end{figure}

\begin{figure}[ht]
    \centering
    \includegraphics[width=0.47\textwidth]{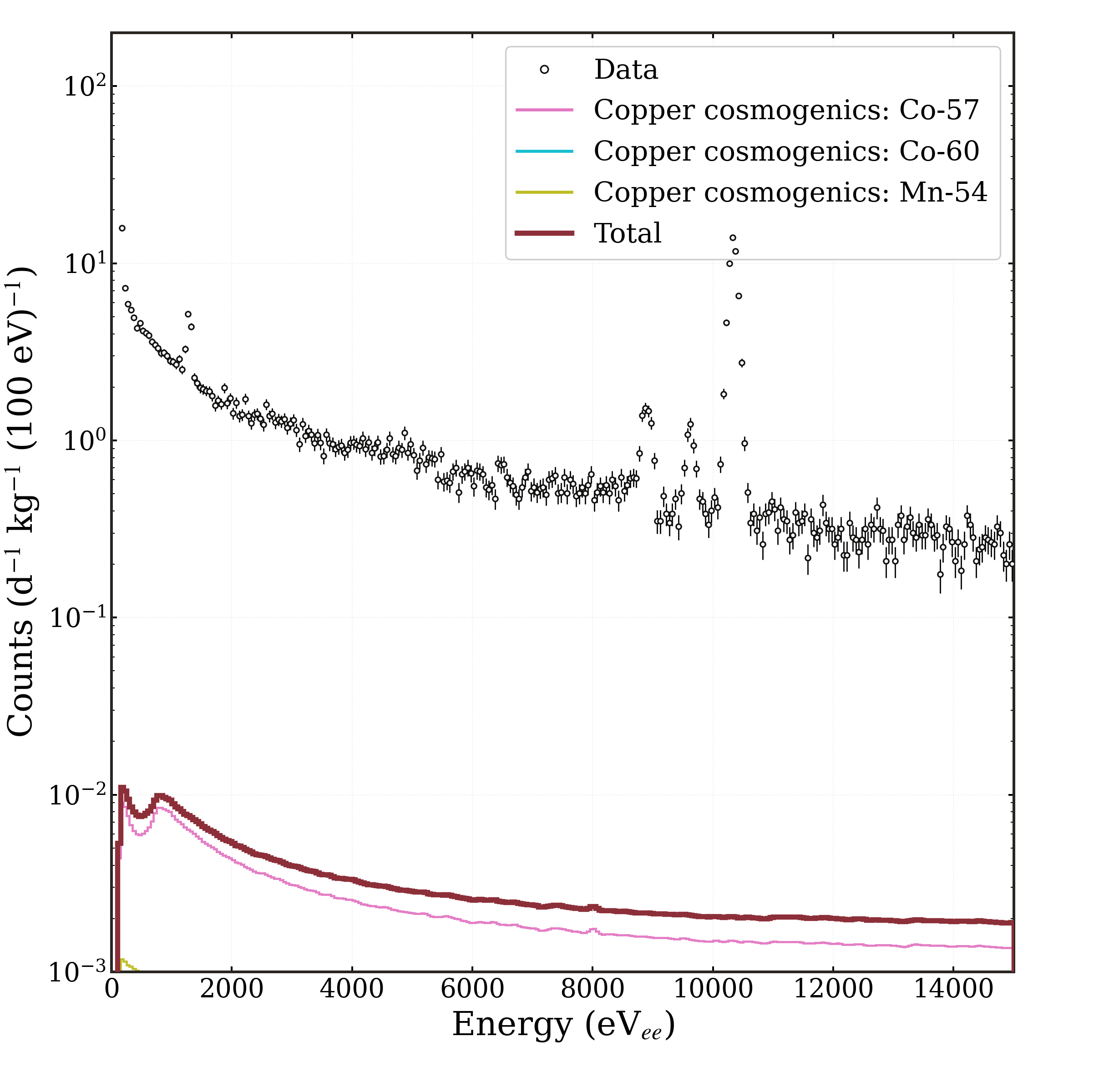}
    \caption{Background contributions from cosmogenically activated isotopes in the copper parts of the cryostat. Their corresponding activities were calculated according to \cite{conus_bkg} assuming an exposure of 100 days. }
    \label{fig:cu_cosm_no_lines}
\end{figure}

All other isotopes either produce no visible lines in the low or high energy channels of the experiment, or their activity is too low. In order to accurately account for them, their production rate as measured previously in the CONUS experiment (see \cite{conus_bkg}) is considered. The time of exposure to cosmic rays for both the germanium crystals and the copper parts is monitored and in the materials were primarily stored in underground facilities whenever possible. A conservative time of (100 $\pm$ 10) d of cosmic ray exposure is assumed in the following. The corresponding MC simulation results for $^{57}$Co, $^{60}$Co, $^{54}$Mn, $^{55}$Fe, and $^{3}$He in the germanium crystals are shown in Figure~\ref{fig:ge_cosm_no_lines}. The results for $^{57}$Co, $^{60}$Co, and $^{54}$Mn in the copper parts of the cryostat are shown in Figure~\ref{fig:cu_cosm_no_lines}. The contributions to the total background of the detectors are small and only account for below 1 $\%$ of the overall count rate in all energy ranges. Thus, the time dependence of the decay is also not relevant for the analysis.

\section{Additional backgrounds}
\label{sec:additional}

Besides the background sources discussed above, three additional contributions are needed for a complete description of the Run~1 spectra: the decay chain of $^{210}$Pb in the innermost lead layer of the shield, a small residual contamination inside the cryostat of C2 and C3, and the low-energy ``leakage test background'' known from the previous CONUS measurement campaign. Their impact on the total spectrum is small compared to cosmic muons and neutrons in the region of interest, but they become relevant once the spectra are decomposed over the full energy range.

\subsection{$^{210}$Pb in lead and cryostat parts}

$^{210}$Pb is a natural daughter isotope of the $^{238}$U chain. For \conusplus, $^{210}$Pb is present in the lead parts of the shield. Only the innermost lead layer needs to be considered explicitly, since the outer layers are efficiently self-shielded. Material screening of the low-activity lead bricks used for the inner layer yielded activities below 1.7~Bq~kg$^{-1}$ at the time of the original CONUS construction. Correcting for the $^{210}$Pb half-life gives an activity below (1.4 $\pm$ 0.1)~Bq~kg$^{-1}$ for the start of Run~1, which is used as input to the MC simulation.

\begin{figure}[ht]
    \centering
    \includegraphics[width=0.47\textwidth]{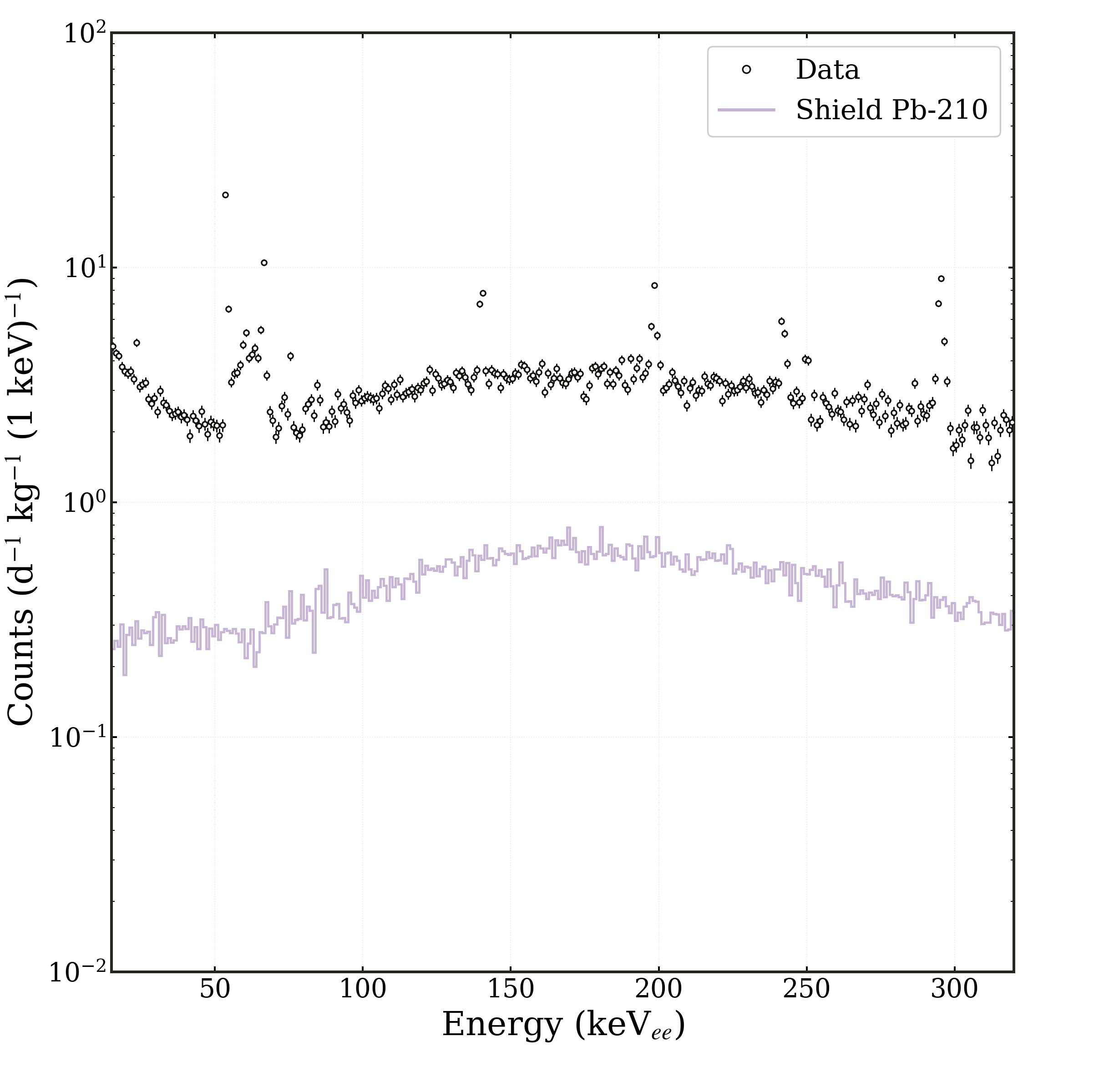}
    \caption{MC simulation result for $^{210}$Pb in the innermost shield layer and comparison to the spectrum of the C5 detector in reactor on measurement in run 1. }
    \label{fig:pb210_shield_result}
\end{figure}

The resulting contribution is small in the low-energy region and negligible in the \CEvNS\ analysis window. It mainly affects the spectrum above a few tens of keV$_{ee}$ and remains well below the dominant background components. A representative MC simulation result is shown in Fig.~\ref{fig:pb210_shield_result}. Contributions from $^{210}$Pb in the cryostat itself are constrained independently from the absence of a visible 46.5~keV line in the data. Several tests were performed by placing $^{210}$Pb contaminations at different positions in the cryostat to test if certain locations could lead to a continuous background without the 46.5~keV line. No such position was found. Any residual $^{210}$Pb contamination inside the cryostat must therefore be at the level of an upper limit and has no relevant impact on the final background model. This shows the improved radiopurity of the cryostats compared to the CONUS experiment, where $^{210}$Pb in the cryostat was found to be one of the largest background contributions \cite{conus_bkg}.  

\subsection{Additional cryostat contamination in C2 and C3}

After including all identified components, an excess remained in the C2 and C3 spectra in both reactor on and reactor off data. The shape of this residual contribution is smooth, extends from the sub-keV region to a few hundred keV, and does not introduce additional gamma-ray lines. This disfavours explanations in terms of a larger external muon or neutron flux and instead points to a weak contamination close to the crystal.

\begin{figure}[ht]
    \centering
    \includegraphics[width=0.47\textwidth]{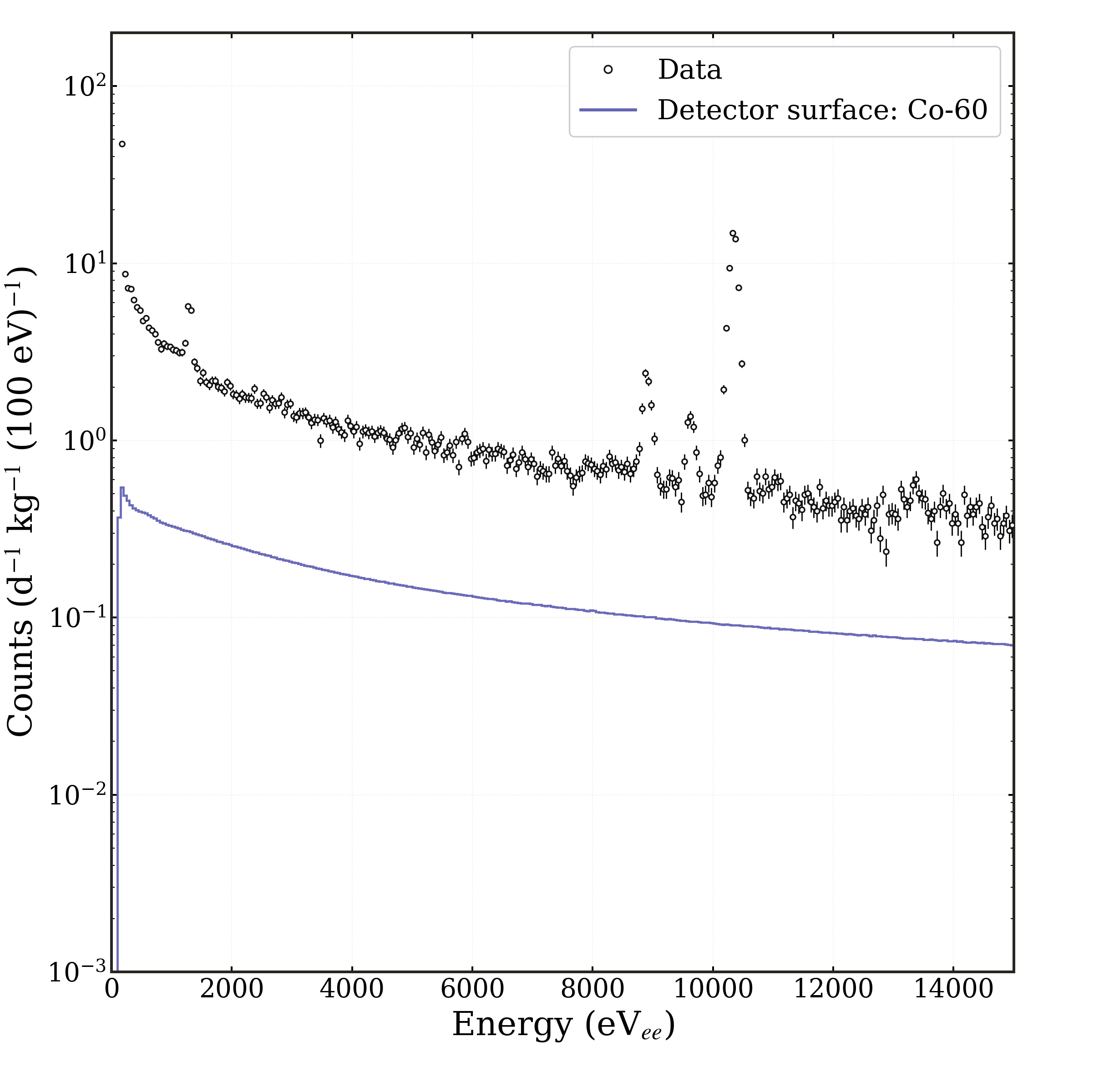}
    \caption{MC simulation result of a 5 $\mu$Bq $^{60}$Co contamination in close proximity to the germanium crystal of C2. The spectrum matches the missing background component in the C2 and C3 background model. }
    \label{fig:co60_cryo_result}
\end{figure}

A satisfactory description is obtained with a small $^{60}$Co activity in copper parts near the passivation-side region of the detectors. Owing to its long half-life, such a contamination can persist from earlier cosmic-ray exposure of individual cryostat parts. An activity of order 5~$\mu$Bq is sufficient to reproduce both the shape and the absolute rate of the missing component. Since this contribution is only required for C2 and C3, it is included detector-by-detector in the final model rather than as a common background term. Figure~\ref{fig:co60_cryo_result} shows the corresponding simulated spectrum.

\subsection{Leakage test background}

A second residual component at low energies is present in C2 and C3 below about 1~keV$_{ee}$. Both detectors were already operated in the predecessor CONUS experiment and were exposed there to a regular containment leakage test during which they were filled with argon gas to avoid any damage to the cryostats \cite{Ackermann_2024}. After the procedure, an additional low-energy background appeared and remained stable over time. The origin of this component could not be linked to a distinct radioactive contamination, but its spectral shape was found to be well described by an exponential increase towards low energies. The same empirical parametrization as in \cite{Ackermann_2024} is therefore retained here,
\begin{equation}
    b_{\mathrm{Leak}}(E) = \theta_1\, e^{-\theta_2 E},
\end{equation}
with detector-dependent normalisation and slope parameters.

In Run~1 of \conusplus, this component is smaller than in the final CONUS campaign, but it is still required for C2 and C3 in order to reproduce the data below 1~keV$_{ee}$. Above a few keV$_{ee}$ its contribution quickly becomes negligible. Since it is present in both reactor states with consistent magnitude, it is treated as a time-independent detector-specific term in the full background model.

\subsection{Impact of reactor-induced high energy gamma-rays and environmental radiation}
\label{sec:highEgammas}

As detailed in Section~\ref{sec:back_char}, the CONUS+ room features high energy $\gamma$ radiation produced from neutron activation of materials in the reactor building. In order to test the possible impact of this radiation on the background of the four detectors, 10$^{10}$ $\gamma$ particles with an energy of 10 MeV were started from a point source on the outside
of the CONUS+ shield in the direction of the centre of the shield in MaGe. No impact in the detectors was found in the MC simulations, indicating the ability to effectively shield all $\gamma$ radiation despite the reduced amount of lead. Due to the high energy which was chosen for the started $\gamma$ particles, any impact of environmental radiation on the CONUS+ measurement can also be excluded.  

\section{Background model}
\label{sec:fullmodel}

\subsection{Reactor on/off differences}

The same set of background components is used to describe reactor on and reactor off data, with only a few well-motivated changes. During the reactor outage, the drywell lid is positioned directly above the experimental room and increases the local overburden. Dedicated MC simulations show that this reduces the cosmic-neutron flux in the room by about 19\% and the muon flux by about 3\%. These corrections are propagated to the corresponding background components in the reactor off model.

A second difference is the radon level inside the detector chamber. The radon flushing system was tuned continuously during the early phase of Run~1, and the reactor off period occurred when the suppression was more effective than during large parts of the preceding reactor on measurement. Consequently, the radon contribution is reduced in the reactor off spectra according to the observed line intensities. Reactor neutrons are absent in reactor off data, while the contribution from short-lived inert gases, in particular $^{135}$Xe, is enhanced immediately after opening the reactor vessel.

\subsection{Combined model and spectral decomposition}

\begin{figure*}[htbp]
  \centering
  \begin{subfigure}[b]{0.95\textwidth}
    \centering
    \includegraphics[width=\textwidth]{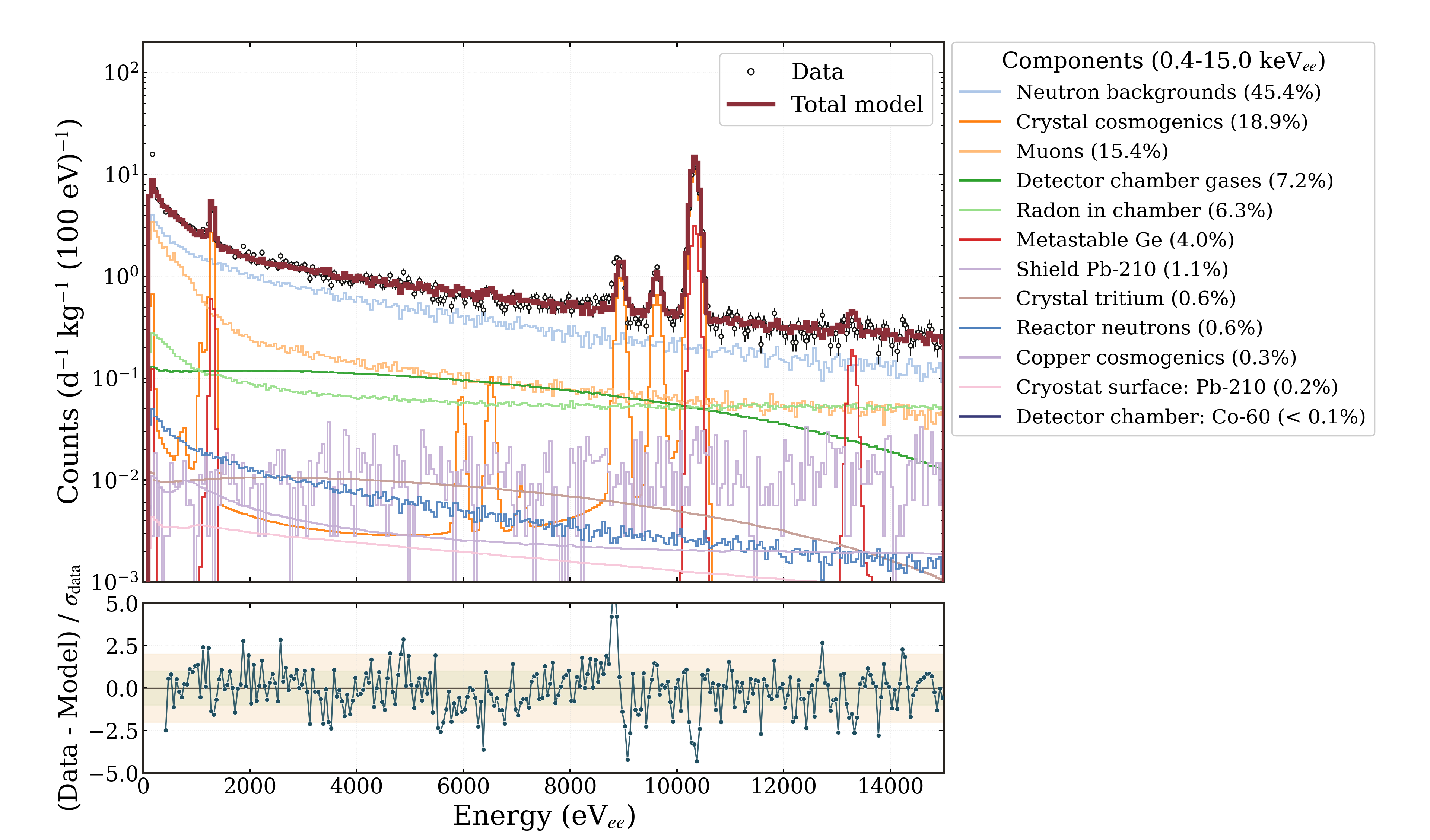}
    \caption{}
    \label{fig:fullmodel_c5_on_low}
  \end{subfigure}
  \hfill
  \begin{subfigure}[b]{0.95\textwidth}
    \centering
    \includegraphics[width=\textwidth]{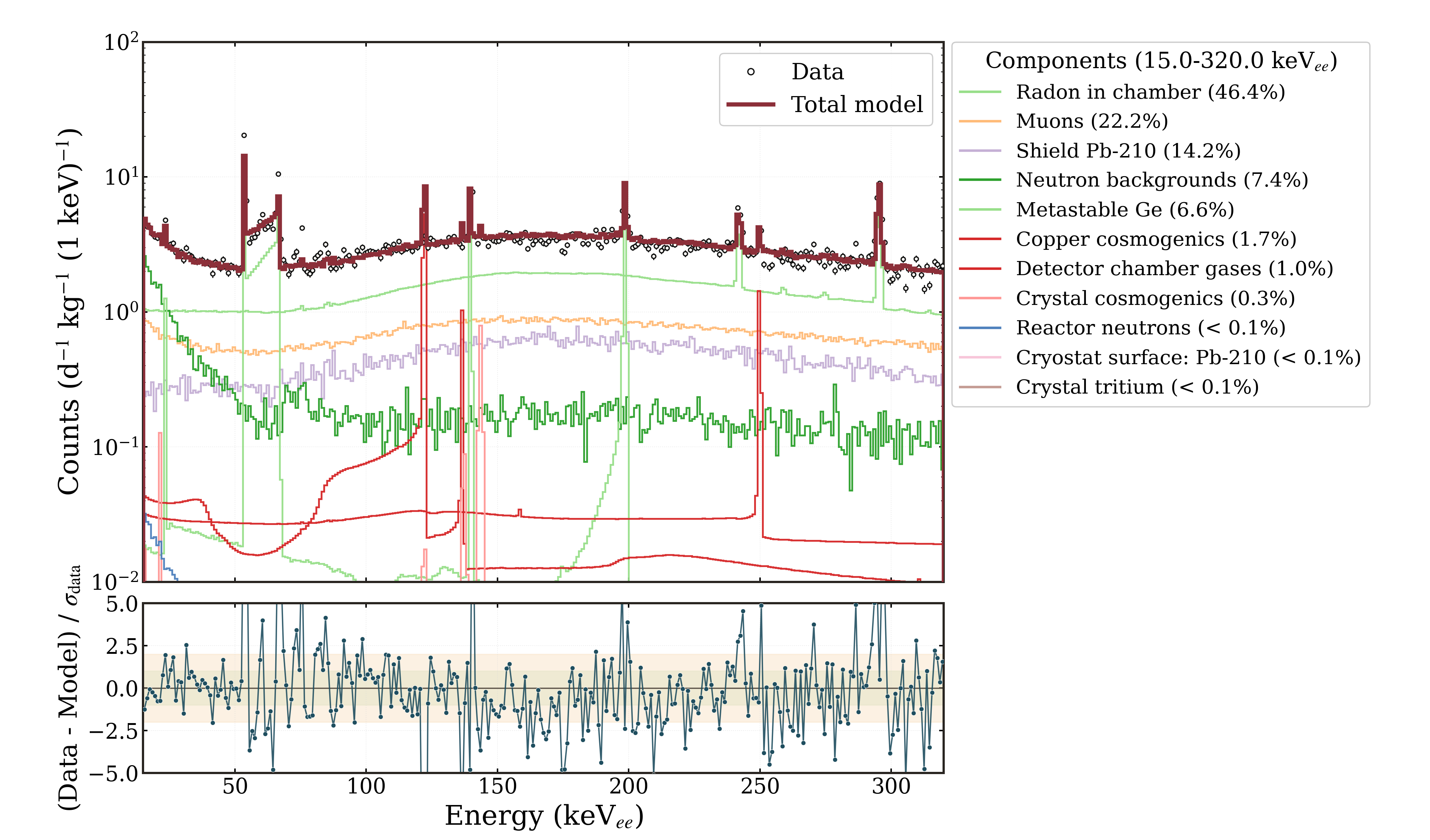}
    \caption{}
    \label{fig:fullmodel_c5_on_high}
  \end{subfigure}
  \caption{Full background model for the C5 detector in reactor on measurement in both low and high energy channels compared to the respective data.}
  \label{fig:fullmodel_c5}
\end{figure*}

\begin{figure*}[htbp]
  \centering
  \begin{subfigure}[b]{0.95\textwidth}
    \centering
    \includegraphics[width=\textwidth]{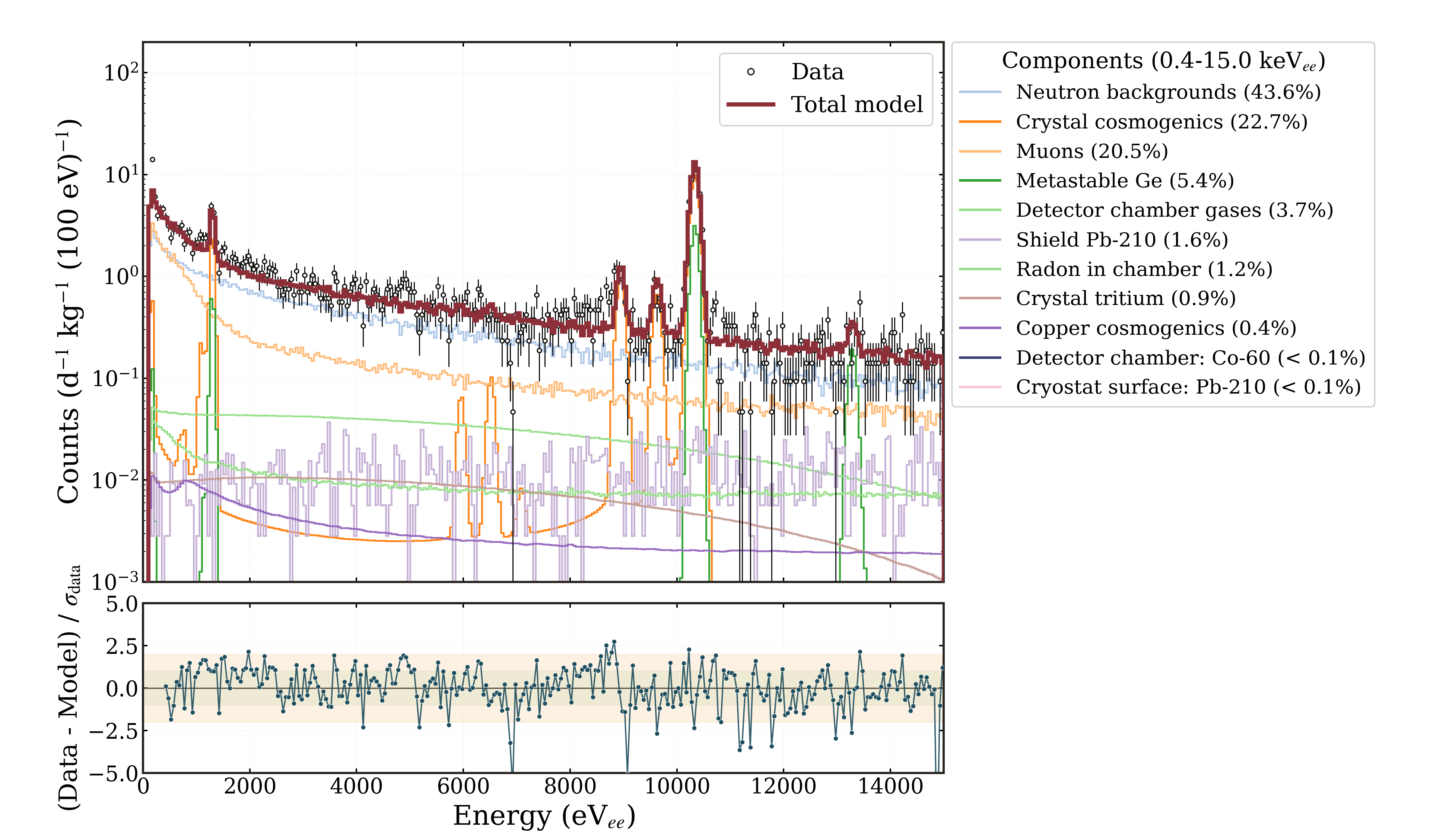}
    \caption{}
    \label{fig:fullmodel_c5_off_low}
  \end{subfigure}
  \hfill
  \begin{subfigure}[b]{0.95\textwidth}
    \centering
    \includegraphics[width=\textwidth]{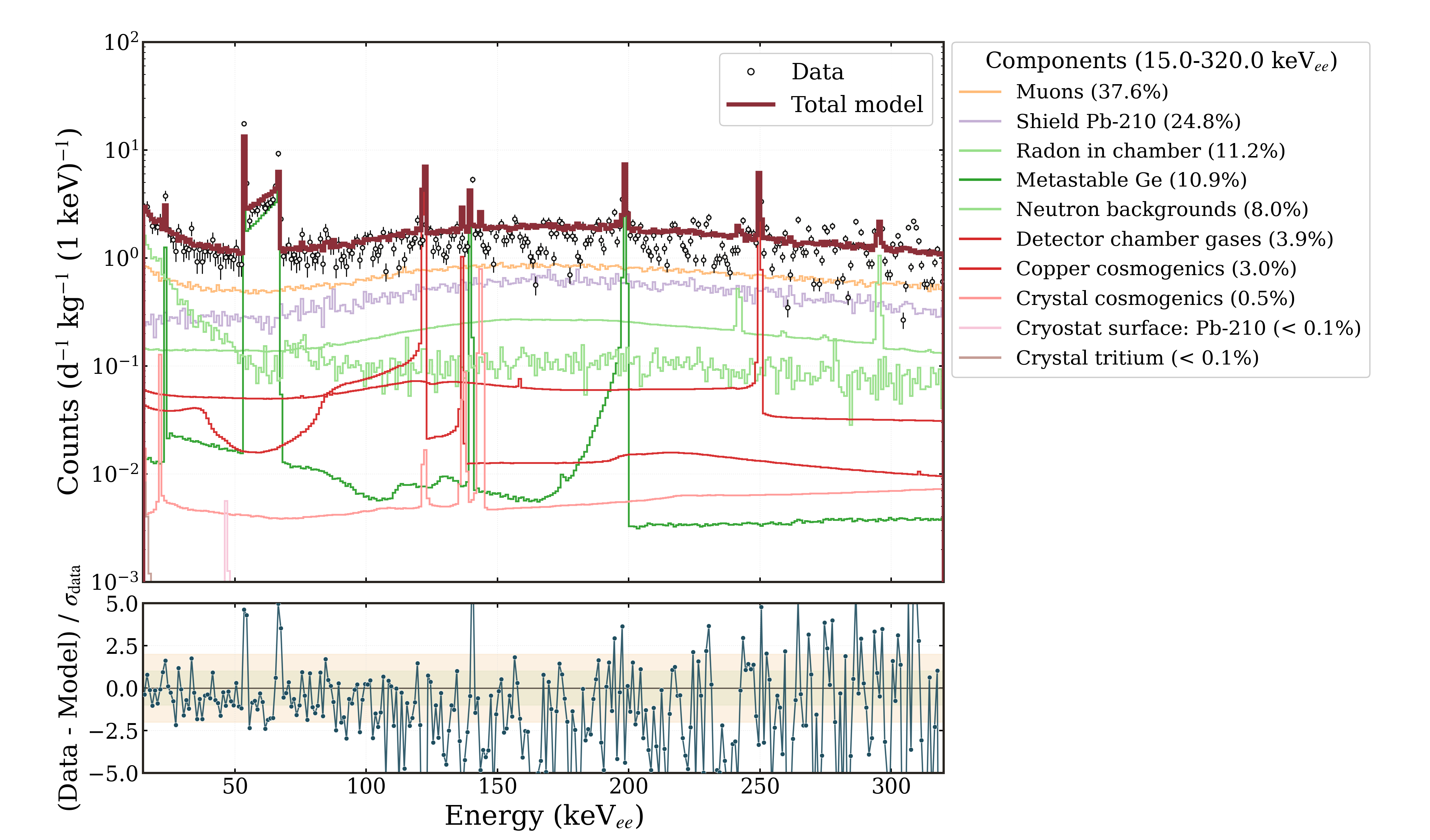}
    \caption{}
    \label{fig:fullmodel_c5_off_high}
  \end{subfigure}
  \caption{Full background model for the C5 detector in reactor off measurement in both low and high energy channels compared to the respective data.}
  \label{fig:fullmodel_c5_off}
\end{figure*}

\begin{figure*}[htbp]
  \centering
  \begin{subfigure}[b]{0.95\textwidth}
    \centering
    \includegraphics[width=\textwidth]{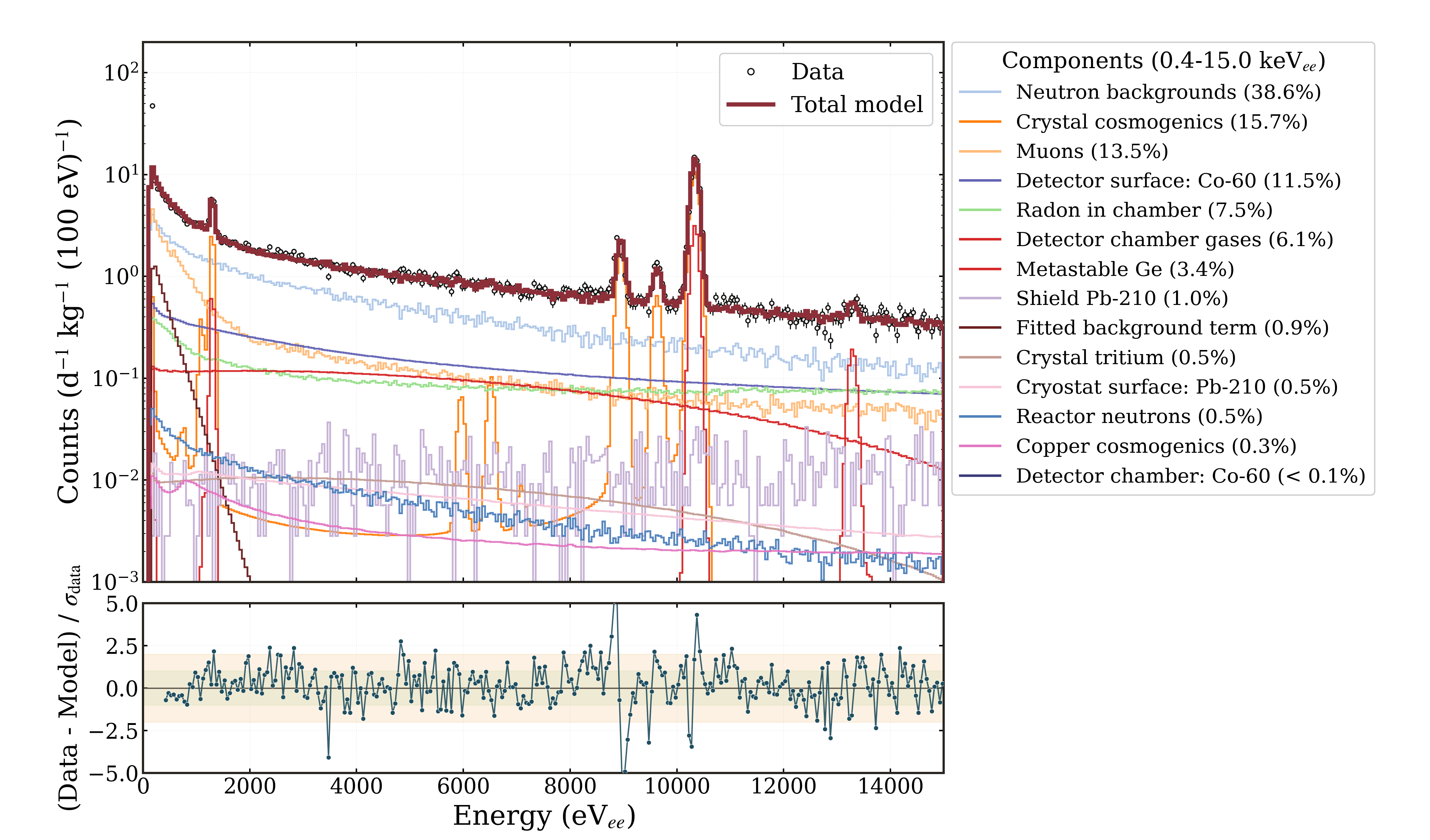}
    \caption{}
    \label{fig:fullmodel_c2_on_low}
  \end{subfigure}
  \hfill
  \begin{subfigure}[b]{0.95\textwidth}
    \centering
    \includegraphics[width=\textwidth]{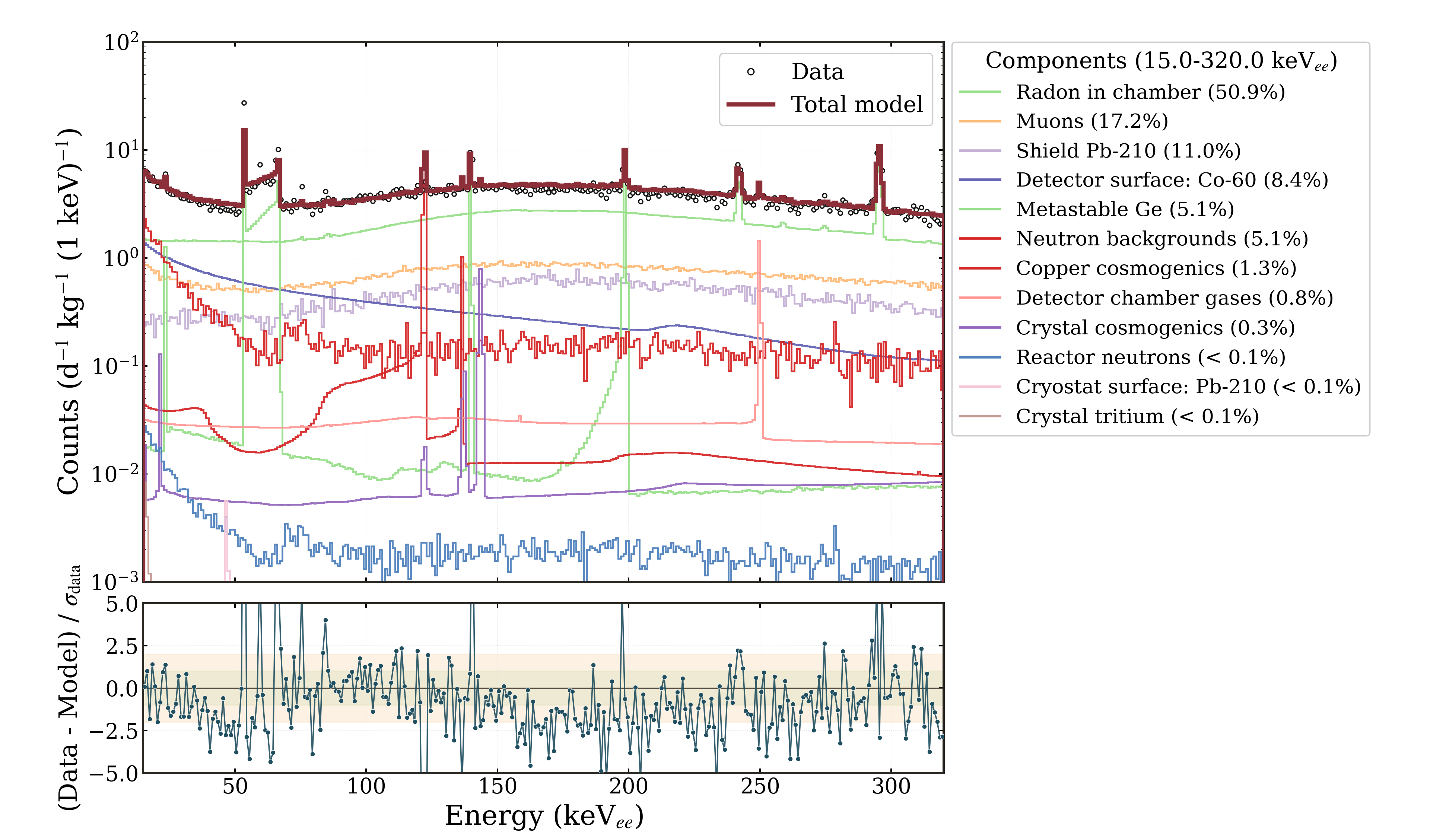}
    \caption{}
    \label{fig:fullmodel_c2_on_high}
  \end{subfigure}
  \caption{Full background model for the C2 detector in reactor on measurement in both low and high energy channels compared to the respective data.}
  \label{fig:fullmodel_c2}
\end{figure*}

\begin{figure*}[htbp]
  \centering
  \begin{subfigure}[b]{0.95\textwidth}
    \centering
    \includegraphics[width=\textwidth]{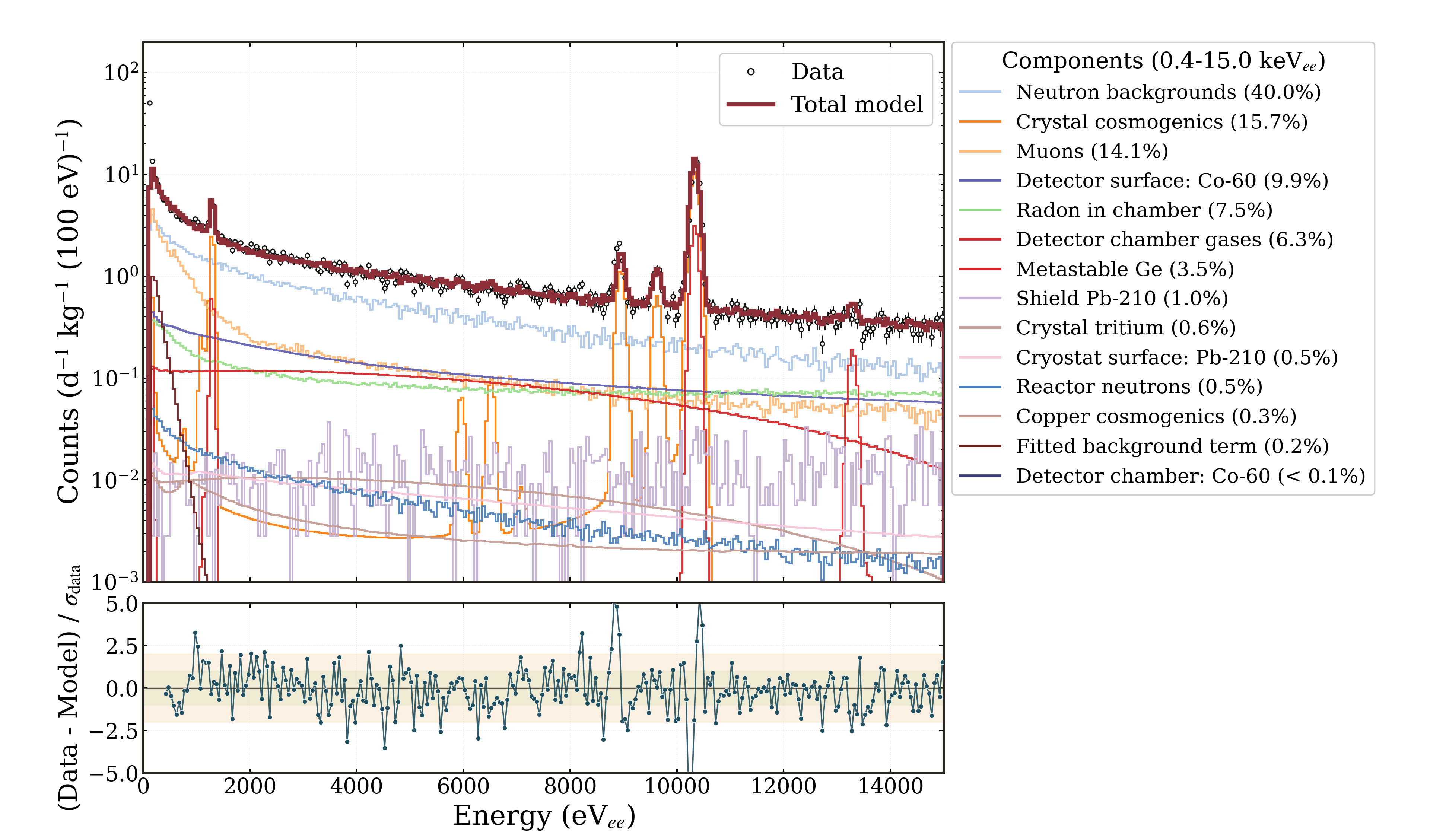}
    \caption{}
    \label{fig:fullmodel_c3_on_low}
  \end{subfigure}
  \hfill
  \begin{subfigure}[b]{0.95\textwidth}
    \centering
    \includegraphics[width=\textwidth]{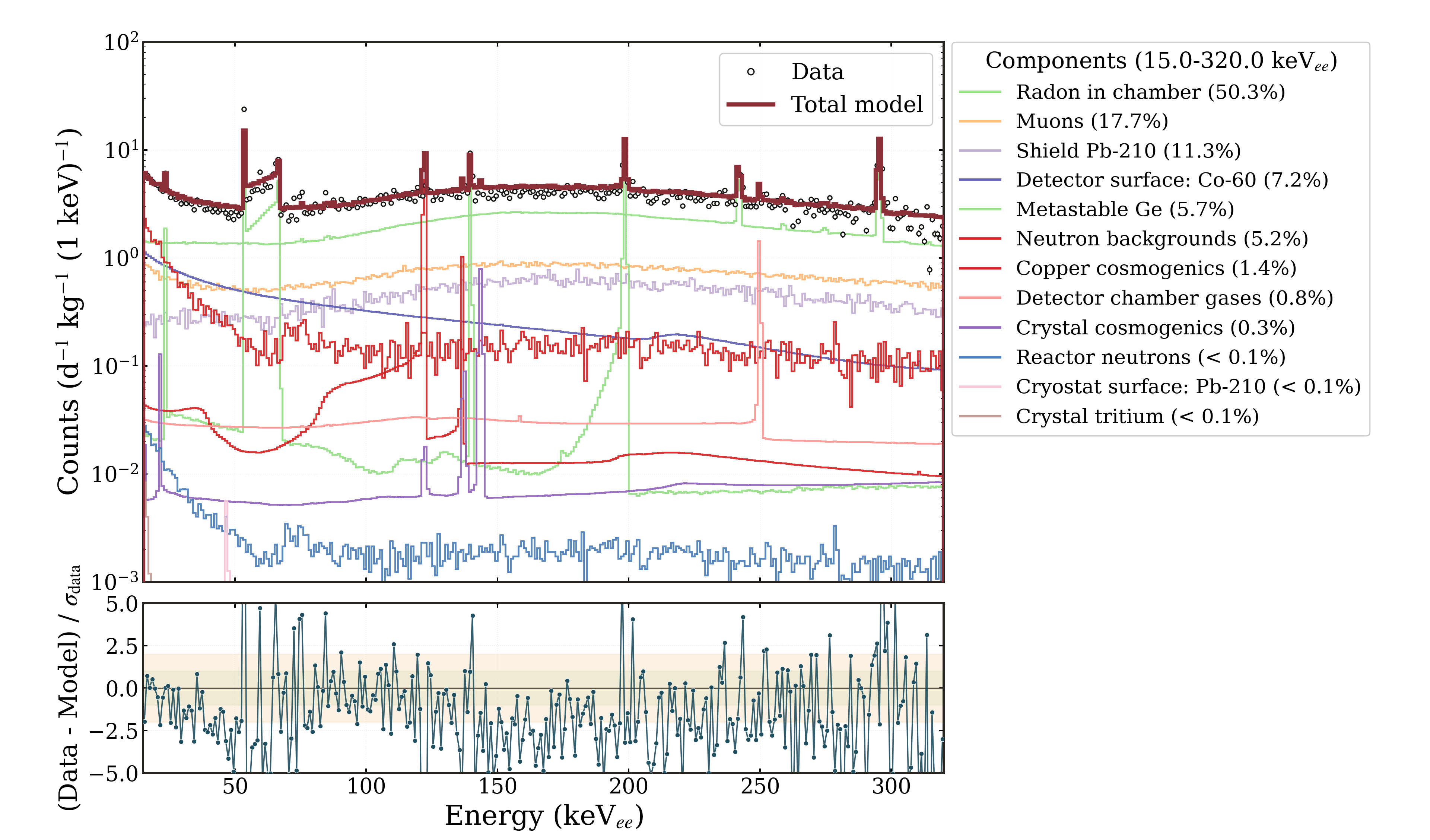}
    \caption{}
    \label{fig:fullmodel_c3_on_high}
  \end{subfigure}
  \caption{Full background model for the C3 detector in reactor on measurement in both low and high energy channels compared to the respective data.}
  \label{fig:fullmodel_c3}
\end{figure*}

The full background model is obtained by summing all simulated and empirical components for each detector and for both reactor states. Figures~\ref{fig:fullmodel_c5} - \ref{fig:fullmodel_c3} show the final comparison of the model to the data of the three detectors in run 1 with their respective decomposition. Overall the model agrees well with the measured data in all energy ranges and measurement periods, with differences within the 1 to 2 $\sigma$ range for most energy ranges. Below 1 keV$_{ee}$, especially good agreement and deviations below 1 $\sigma$ are found. In the \CEvNS\ region of interest, the spectra are dominated by cosmic-ray induced backgrounds, namely the direct muon component together with cosmic neutrons and muon-induced neutrons from the overburden. Depending on detector and reactor state, these components account for roughly 70--80\% of the total background between 0.4 and 1~keV$_{ee}$.

At higher energies the composition changes. The relative importance of cosmic neutrons decreases, while electromagnetic energy deposition from muons remains visible. Radon becomes one of the leading contributions above several tens of keV$_{ee}$, especially in reactor on data, and the remaining cosmogenic and $^{210}$Pb-related terms account for smaller but non-negligible fractions. The additional detector-specific components discussed in Section~\ref{sec:additional} are required mainly for C2 and C3 and improve the agreement in the sub-keV region and in the broad continuum up to a few hundred keV$_{ee}$. A compact overview of the decomposition for all three detectors and both reactor states is given in Table~\ref{tab:decomposition_summary}.

\begin{table*}[t]
\centering
\footnotesize
\setlength{\tabcolsep}{3pt}
\renewcommand{\arraystretch}{0.96}
\caption{Overview (summary) of the Run~1 background decomposition for C5, C2, and C3. For each entry, values are given as reactor off / reactor on. A dash indicates that the corresponding reactor-on interval below 0.4~keV$_{ee}$ was blinded. The first energy interval extends from the detector threshold to 0.4~keV$_{ee}$; the thresholds are detector-dependent and correspond to 0.17~keV$_{ee}$ for C5 off, 0.18~keV$_{ee}$ for C2 off, 0.18~keV$_{ee}$ for C3 off, 0.16~keV$_{ee}$ for C5 on, 0.18~keV$_{ee}$ for C2 on, and 0.16~keV$_{ee}$ for C3 on. Percentages denote the relative contribution to the total background model in the respective interval.}
\label{tab:decomposition_summary}
\resizebox{\textwidth}{!}{%
\begin{tabular}{lcccccc}
\hline
 & $[\mathrm{thr},0.4]$~keV$_{ee}$ & $[0.4,1]$~keV$_{ee}$ & $[2,8]$~keV$_{ee}$ & $[15,30]$~keV$_{ee}$ & $[30,100]$~keV$_{ee}$ & $[100,250]$~keV$_{ee}$ \\
\hline
\multicolumn{7}{c}{\textbf{C5}} \\
\hline
Data [d$^{-1}$ kg$^{-1}$] & $23.5\pm1.1$ / -- & $33.4\pm1.3$ / $42.9\pm0.6$ & $74.8\pm2.0$ / $103.6\pm0.9$ & $30.7\pm1.3$ / $51.4\pm0.7$ & $133.0\pm2.7$ / $220.9\pm1.4$ & $242.6\pm3.6$ / $512.5\pm2.1$ \\
Bkg. model [d$^{-1}$ kg$^{-1}$] & $23.6\pm1.5$ / -- & $35.8\pm2.9$ / $42.9\pm3.1$ & $75.4\pm6.4$ / $103.8\pm8.3$ & $32.3\pm1.6$ / $48.6\pm2.2$ & $135.8\pm3.2$ / $204.5\pm4.6$ & $283.7\pm4.8$ / $529.6\pm9.1$ \\
Cosmic neutrons & 44.6\% / -- & 52.9\% / 50.3\% & 62.5\% / 56.4\% & 41.0\% / 30.6\% & 8.3\% / 7.1\% & 6.6\% / 4.5\% \\
Reactor neutrons & -- / -- & -- / 0.7\% & -- / 0.7\% & -- / 0.4\% & -- / 0.1\% & -- / 0.1\% \\
$\mu$-induced n (overburden) & 5.0\% / -- & 5.5\% / 5.2\% & 6.6\% / 6.0\% & 4.4\% / 3.3\% & 0.8\% / 0.6\% & 0.7\% / 0.4\% \\
Cosmic muons & 46.3\% / -- & 45.4\% / 35.5\% & 20.6\% / 15.3\% & 31.5\% / 19.5\% & 28.2\% / 17.6\% & 48.9\% / 23.9\% \\
Cu cosmogenics & 0.1\% / -- & 0.3\% / 0.3\% & 0.5\% / 0.4\% & 1.7\% / 1.0\% & 1.8\% / 1.1\% & 5.3\% / 2.5\% \\
Ge cosmogenics & 1.7\% / -- & 0.7\% / 0.2\% & 0.4\% / 0.4\% & 0.2\% / 0.1\% & 0.2\% / 0.1\% & 0.3\% / 0.2\% \\
Metastable Ge states & 0.2\% / -- & $<0.1$\% / $<0.1$\% & 0.1\% / 0.1\% & 4.6\% / 2.9\% & 39.0\% / 21.7\% & 4.4\% / 2.7\% \\
$^{210}$Pb (shield + cryostat) & 0.1\% / -- & 0.4\% / 0.3\% & 1.8\% / 1.6\% & 12.2\% / 7.6\% & 16.4\% / 9.9\% & 34.6\% / 16.4\% \\
Radon & 0.7\% / -- & 0.8\% / 4.5\% & 1.4\% / 7.5\% & 6.5\% / 28.1\% & 7.7\% / 33.6\% & 15.1\% / 51.7\% \\
Inert gases from reactor & 1.0\% / -- & 1.7\% / 3.3\% & 6.0\% / 11.9\% & 2.7\% / 1.4\% & 2.8\% / 0.9\% & 5.9\% / 1.2\% \\
\hline
\multicolumn{7}{c}{\textbf{C2}} \\
\hline
Data [d$^{-1}$ kg$^{-1}$] & $30.6\pm1.3$ / -- & $45.3\pm1.5$ / $52.3\pm0.7$ & $105.4\pm2.4$ / $130.6\pm1.0$ & $52.2\pm1.7$ / $71.0\pm0.8$ & $201.0\pm3.3$ / $281.3\pm1.5$ & $386.0\pm4.5$ / $643.2\pm2.3$ \\
Bkg. model [d$^{-1}$ kg$^{-1}$] & $30.8\pm2.1$ / -- & $45.2\pm3.3$ / $50.4\pm3.1$ & $97.5\pm6.8$ / $127.3\pm8.6$ & $51.2\pm2.3$ / $70.5\pm2.6$ & $189.7\pm4.8$ / $272.7\pm5.5$ & $388.3\pm6.4$ / $678.2\pm12.1$ \\
Cosmic neutrons & 32.5\% / -- & 39.0\% / 42.8\% & 44.4\% / 44.8\% & 24.1\% / 22.2\% & 5.5\% / 4.9\% & 4.1\% / 3.2\% \\
Reactor neutrons & -- / -- & -- / 0.6\% & -- / 0.6\% & -- / 0.3\% & -- / 0.1\% & -- / 0.1\% \\
$\mu$-induced n (overburden) & 3.6\% / -- & 4.0\% / 4.4\% & 4.7\% / 4.7\% & 2.6\% / 2.4\% & 0.6\% / 0.5\% & 0.5\% / 0.3\% \\
Cosmic muons & 40.1\% / -- & 36.2\% / 32.8\% & 13.8\% / 12.3\% & 18.5\% / 14.1\% & 19.1\% / 13.8\% & 31.6\% / 18.9\% \\
Cu cosmogenics & 0.1\% / -- & 0.2\% / 0.2\% & 0.4\% / 0.3\% & 1.0\% / 0.7\% & 1.2\% / 0.9\% & 3.3\% / 2.0\% \\
Ge cosmogenics & 1.0\% / -- & 0.5\% / 0.3\% & 0.8\% / 0.7\% & 0.1\% / 0.1\% & 0.2\% / 0.1\% & 0.2\% / 0.2\% \\
Metastable Ge states & 0.1\% / -- & $<0.1$\% / $<0.1$\% & 0.1\% / 0.1\% & 2.7\% / 2.0\% & 23.9\% / 17.2\% & 3.6\% / 2.2\% \\
$^{210}$Pb (shield + cryostat) & 0.1\% / -- & 0.2\% / 0.5\% & 1.3\% / 1.7\% & 7.2\% / 6.1\% & 10.9\% / 7.8\% & 21.7\% / 13.0\% \\
Radon & 1.2\% / -- & 1.5\% / 5.5\% & 2.5\% / 8.5\% & 9.4\% / 28.9\% & 12.5\% / 37.5\% & 23.3\% / 58.4\% \\
Inert gases from reactor & 0.7\% / -- & 1.2\% / 2.8\% & 4.3\% / 9.5\% & 1.5\% / 1.0\% & 1.0\% / 0.9\% & 1.6\% / 1.0\% \\
$^{60}$Co in cryostat & 6.5\% / -- & 9.9\% / 8.7\% & 18.4\% / 14.8\% & 29.6\% / 21.8\% & 19.4\% / 13.9\% & 10.5\% / 6.3\% \\
Leakage test background & 14.3\% / -- & 6.5\% / 5.6\% & 0.1\% / 0.1\% & -- / -- & -- / -- & -- / -- \\
\hline
\multicolumn{7}{c}{\textbf{C3}} \\
\hline
Data [d$^{-1}$ kg$^{-1}$] & $31.1\pm1.3$ / -- & $42.5\pm1.5$ / $48.8\pm0.6$ & $98.4\pm2.3$ / $123.3\pm1.0$ & $49.6\pm1.6$ / $66.7\pm0.8$ & $182.6\pm3.1$ / $261.7\pm1.5$ & $387.6\pm4.5$ / $606.6\pm2.3$ \\
Bkg. model [d$^{-1}$ kg$^{-1}$] & $30.9\pm2.1$ / -- & $42.0\pm3.2$ / $49.2\pm3.5$ & $93.2\pm7.4$ / $123.4\pm8.4$ & $48.0\pm2.2$ / $66.9\pm2.6$ & $182.9\pm4.0$ / $256.4\pm5.2$ & $381.2\pm5.9$ / $658.9\pm11.3$ \\
Cosmic neutrons & 33.7\% / -- & 41.6\% / 44.2\% & 47.6\% / 47.4\% & 25.4\% / 23.6\% & 6.0\% / 5.4\% & 4.2\% / 3.3\% \\
Reactor neutrons & -- / -- & -- / 0.6\% & -- / 0.6\% & -- / 0.3\% & -- / 0.1\% & -- / 0.1\% \\
$\mu$-induced n (overburden) & 3.8\% / -- & 4.3\% / 4.6\% & 5.0\% / 5.0\% & 2.7\% / 2.6\% & 0.7\% / 0.5\% & 0.6\% / 0.4\% \\
Cosmic muons & 42.6\% / -- & 37.7\% / 33.8\% & 15.8\% / 13.0\% & 19.5\% / 15.0\% & 21.2\% / 15.1\% & 31.5\% / 20.0\% \\
Cu cosmogenics & 0.1\% / -- & 0.2\% / 0.2\% & 0.4\% / 0.3\% & 1.1\% / 0.8\% & 1.3\% / 1.0\% & 3.3\% / 2.1\% \\
Ge cosmogenics & 1.4\% / -- & 0.5\% / 0.4\% & 0.4\% / 0.8\% & 0.1\% / 0.1\% & 0.2\% / 0.1\% & 0.2\% / 0.2\% \\
Metastable Ge states & 0.2\% / -- & $<0.1$\% / $<0.1$\% & 0.1\% / 0.1\% & 2.9\% / 2.2\% & 26.2\% / 18.9\% & 3.6\% / 3.0\% \\
$^{210}$Pb (shield + cryostat) & 0.1\% / -- & 0.2\% / 0.5\% & 1.4\% / 1.8\% & 7.5\% / 6.4\% & 11.9\% / 8.5\% & 21.6\% / 13.8\% \\
Radon & 1.2\% / -- & 1.6\% / 5.4\% & 2.7\% / 8.6\% & 9.9\% / 29.4\% & 13.8\% / 39.3\% & 23.2\% / 59.3\% \\
Inert gases from reactor & 0.7\% / -- & 1.3\% / 2.9\% & 4.6\% / 10.0\% & 1.7\% / 1.0\% & 1.1\% / 0.8\% & 1.6\% / 1.0\% \\
$^{60}$Co in cryostat & 5.6\% / -- & 8.7\% / 7.4\% & 16.2\% / 13.0\% & 25.8\% / 19.2\% & 17.6\% / 12.5\% & 8.6\% / 5.5\% \\
Leakage test background & 9.8\% / -- & 1.9\% / 1.5\% & 0.1\% / 0.1\% & -- / -- & -- / -- & -- / -- \\
\hline
\end{tabular}%
}
\end{table*}

\subsection{Systematic and statistical uncertainties of the model}
\label{sec:sys}

Because of the important role of the background model in the final likelihood fit of the CONUS+ run 1 data, the uncertainty of the model is an important part of the total uncertainty of the CE$\nu$NS result. The overall uncertainty of the total model in the different energy regions is shown in Table~\ref{tab:decomposition_summary}. For each specific component of the model, uncertainties mainly come from the different  normalisation methods and will be explained in the following section. \newline
In the case of the muonic contribution, the  normalisation directly stems from the flux which is calculated from muon flux models from literature \cite{reyna, Bugaev}. These calculations are mainly based on the effective overburden of the experiment, which was measured to be (7.3 $\pm$ 0.1) m w.e. which yields an uncertainty of $\pm$6 $\%$ on the flux. The uncertainty directly translates to the final uncertainty on the muon background count rate. The uncertainty of the neutron background count rate of approximately 15 \% mainly comes from the original assumed uncertainty for the outside neutron flux which was taken from literature. The uncertainty of the overburden also has an effect on this uncertainty, but it is subdominant compared to the initial flux which was tested with a dedicated MC simulation. As the neutron component currently features the largest uncertainty in the model while also being the most dominant contribution at low energies, the CONUS+ collaboration is currently planning a dedicated neutron measurement, which will be able to verify these MC simulation results independently and reduce the uncertainty. \newline 
Uncertainties for components with specific gamma lines in the low or high energy data like radon or the cosmogenic germanium isotopes stem from the statistical uncertainty of the count rates in the peaks which were used to scale the MC simulation results. These uncertainties are small compared to the muon and neutron contributions in the order of 2 - 3 $\%$. Furthermore, since the corresponding components only contribute to a small part of the total background rate at low energies, their overall significance for the CONUS CE$\nu$NS analysis is small. For cosmogenically induced isotopes in the copper and germanium without any visible lines, the uncertainty on their background rate contribution comes from the uncertainty on the cosmic ray exposure time of the specific material, which was taken to be $\pm$10 $\%$. This directly scales linearly to the final uncertainty of the rate of the specific component. As such, the relative uncertainty of these components is large, however the impact on the overall model is again small due to the low total rate from these background sources. \newline 
For $^{210}$Pb in the innermost layer of the lead shield, uncertainties are derived from the activity of $^{210}$Pb in the layer, which is 14 $\%$. Lastly, the uncertainty on the leakage test background which is found in the C2 and C3 detectors assumes a conservative value of 10 $\%$. As this is the only component of the model, which does not correspond to a known physical background, the uncertainty was estimated based on the shape of the missing background in the data sets of both detectors. The overall background rate from this components is again small and the larger uncertainty therefore has only limited impact on the overall model. \newline

\section{Summary and conclusions}
\label{conclusions}

A full decomposition of the Run~1 background of the CONUS+ experiment was developed on the basis of dedicated characterization measurements and Monte Carlo simulations. The dominant contribution in the \CEvNS\ region of interest is due to cosmic rays. Direct muon interactions (40 - 45 $\%$), cosmic neutrons (30 - 45 $\%$), and muon-induced neutrons in the overburden  (3 - 5 $\%$) together account for the largest fraction of the count rate below 1~keV$_{ee}$, while reactor-induced backgrounds remain subdominant in all energy regions and are suppressed by one order of magnitude compared to the expected CE$\nu$NS signal. This validates the result of \cite{ackermann2025observationreactorantineutrinoscoherent} and excludes the possibility of a misidentification of reactor neutrons as neutrinos. 

At higher energies, radon in the detector chamber and a number of smaller material-related contributions become increasingly important. Here it was shown that radon and $^{210}$Pb in the innermost lead layer account for up to 65 $\%$ of the count rate between 100 and 250 keV$_{ee}$ in the measured reactor on data. All visible lines in the data are explained by cosmogenically induced isotopes in the germanium crystals or radon, with the dominant X-Ray lines at 10.4 keV and below coming from neutron-induced germanium isotopes. The overall influence of these cosmogenics and general material contaminations is very small in the overall model. 

The observed differences between reactor on and reactor off data are quantitatively explained by the changed overburden from the drywell lid, the reduced radon level during the outage period, the disappearance of reactor neutrons, and the transient enhancement from short-lived inert gases.

With respect to the background model which was developed for the predecessor experiment CONUS in \cite{conus_bkg}, several differences can be identified. First, the CONUS+ detectors show no $^{210}$Pb contamination inside the cryostat endcap, which was a dominant background source for CONUS. This change was made possible by avoiding specific materials like lead-based solder in the construction of the detectors. Additionally, the CONUS+ background model is dominated by cosmic neutrons at low energies, a background source which was irrelevant for the CONUS experiment due to its larger overburden of 24 m w.e. Muon-induced backgrounds, the other dominant background source at low energies in the presented model, also had a similar impact in CONUS. 

The CONUS+ background model reproduces the measured spectra of all analysed detectors in both reactor states over the full relevant energy range. Agreement is especially good for all energies below 30 keV$_{ee}$ and specifically below 1 keV$_{ee}$, where the ROI for the CE$\nu$NS analysis of CONUS+ is located. This provides the basis for the likelihood analysis of the run~1 data and for the extraction of the reactor \CEvNS\ signal.  \newline 
Since the end of run 1, the CONUS+ experiment has entered its second measurement campaign, called run 2. In this new campaign, the experimental setup has changed and three of the four 1 kg HPGe detectors were exchanged for larger 2.4 kg HPGe detectors. The change more than doubles the active mass of the experiment and allows for high precision CE$\nu$NS measurements in the future. With these new detectors and the upcoming analysis of their data also comes the need for the development of a new background model. It is generally expected that the overall composition of the model will stay the same, as the dominant components of cosmic muons and neutrons will stay dominant in the new detectors and the materials of the new detectors have been screened before deployment. Nevertheless, new background contributions are possible and will be investigated for the analysis. Additionally, the possibility of pulse shape discrimination as previously used in run 5 of the CONUS experiment \cite{conus_psd} is being investigated for the upcoming data set. The study of the pulse shapes and rejection of near-surface events in this approach will give further insight into the background model and can be used to validate not only the upcoming run 2 model but also the approach for the run 1 model.


\bibliographystyle{bibliostyle}
\bibliography{references}

\end{document}